\documentclass[journal]{IEEEtran}
\usepackage{cite}

\ifCLASSINFOpdf
  \usepackage[pdftex]{graphicx}
  \graphicspath{{./images/pdf/}{./images/jpeg/}}
  \DeclareGraphicsExtensions{.pdf,.png}
\else
\fi
\usepackage{bm}
\usepackage{amsmath}
\usepackage{amssymb}
\usepackage{cases}
\usepackage{soul,color,xcolor}
\newtheorem{theorem}{Theorem}[section]

\newtheorem{corollary}[theorem]{Corollary}

\newtheorem{remark}{Remark}
\newcommand{\tr}[1]{#1^{\!\mathsf{T}}}
\newcommand\inv[1]{#1^{\!-\!1}}
\newcommand{\stcomp}[1]{{#1}^{\mathsf{c}}}
\newcommand{\supp}[1]{\mathrm{supp}(#1)}
\newcommand{\opt}[1]{\widehat{#1}}
\newcommand{\mat}[1]{\mathbf{#1}}

\definecolor{darkgreen}{RGB}{0,128,0}

\usepackage{array}
\newcolumntype{C}[1]{>{\centering}p{#1}}
\usepackage{multirow}

\begin{document}
%
\title{Perfect Recovery Conditions \\ For Non-Negative Sparse Modeling}
%
%
%

\author{Yuki~Itoh,~\IEEEmembership{Student~Member,~IEEE,}
       Marco~F.~Duarte,~\IEEEmembership{Senior~Member,~IEEE,}
       and~Mario~Parente,~\IEEEmembership{Senior~Member,~IEEE}
\thanks{\textcopyright\hspace{4pt}2016 IEEE. Personal use of this material is permitted. However, permission to use this material for any other purposes must be obtained from the IEEE by sending a request to pubs-permissions@ieee.org.}
\thanks{This work was supported in part by the National Science Foundation under grant number IIS-1319585.}
\thanks{The authors are with the Department of Electrical and Computer Engineering, University of Massachusetts, Amherst, MA 01003, U.S.A. (e-mail: yitoh@umass.edu; \{mduarte, mparente\}@ecs.umass.edu).}%
\thanks{An early version of this work appeared in the IEEE Geoscience and Remote Sensing Workshop on Hyperspectral Signal Processing: Evolution in Remote Sensing (WHISPERS), 2015~\cite{Yuki2015Whispers}.}}

\maketitle

\begin{abstract}
Sparse modeling has been widely and successfully used in many applications such as computer vision, machine learning, and pattern recognition. Accompanied with those applications, significant research has studied the theoretical limits and algorithm design for convex relaxations in sparse modeling. However, theoretical analyses on non-negative versions of sparse modeling are limited in the literature either to a noiseless setting or a scenario with a specific statistical noise model such as Gaussian noise. 
This paper studies the performance of non-negative sparse modeling in a more general scenario where the observed signals have an unknown arbitrary distortion, especially focusing on non-negativity constrained and L1-penalized least squares, and gives an exact bound for which this problem can recover the correct signal elements. We pose two conditions to guarantee the correct signal recovery: minimum coefficient condition (MCC) and nonlinearity vs.\ subset coherence condition (NSCC). The former defines the minimum weight for each of the correct atoms present in the signal and the latter defines the tolerable deviation from the linear model relative to the positive subset coherence (PSC), a novel type of ``coherence'' metric. We provide rigorous performance guarantees based on these conditions and experimentally verify their precise predictive power in a hyperspectral data unmixing application. 
\end{abstract}


%
\IEEEpeerreviewmaketitle

\section{Introduction}
\label{sec:introduction}
%
%
%
%
\IEEEPARstart{S}{parse} modeling has achieved significant recognition in a variety of research areas such as signal processing, machine learning, computer vision, and pattern recognition. Sparse models refer to the formulation of a signal of interest (or an approximation of it) as the linear combination of a small number of elements (known as atoms) drawn from a so-called sparsity dictionary (or dictionary for short). The {\em sparse recovery} problem refers to the identification of the relevant dictionary atoms for a particular signal of interest.
Sparse modeling and recovery has a rich history in signal processing, and has received significant attention recently due to the emergence of compressed sensing~\cite{Donoho2006,CandesRIP}, a framework for compressed signal acquisition that leverages sparse modeling.

We can further restrict the coefficients of the atoms to be non-negative. In general, non-negativity is advantageous as it makes the model parameters more interpretable. For instance, Lee and Seung present non-negative matrix factorization~\cite{Lee1999}, which can learn a part-based representation of faces or documents. Just adding non-negativity constraints on a linear model to decompose spectral data gives the model coefficients the meaning of fractional abundances~\cite{Heinz2001}. Non-negative constraints have been applied to independent component analysis in face recognition tasks\cite{plumbley2003algorithms}. 

Many approaches have combined non-negativity and sparse modeling. By adding non-negative constraints, several researchers~\cite{Ji2009ICMLA,He2013TNNLS} refined the performance of applying sparse modeling on a face recognition task obtained by Wright et al.~\cite{Wright2009a}. 
Non-negative least squares (NNLS) has been traditionally used, sometimes accompanied with abundance sum-to-one constraints, to extract the spectral components from hyperspectral pixels (e.g.,~\cite{Heinz2001}), a process called {\em spectral unmixing}. Recently, NNLS has been combined with sparsity with improvements in the unmixing performance~\cite{Szlam2010ICIP,Iordache2011TGRS}. Other examples of combining non-negativity and sparse modeling can be found in astronomical imaging~\cite{Bardsley2006JMAA}, proteomics~\cite{Slawski2014}, and economics~\cite{Wu2014}. It has been noted that sparse solutions can be obtained by NNLS with subsequent thresholding~\cite{Slawski2011,Foucart2014,Slawski2013}.


Several contributions on theoretical analysis of non-negative sparse modeling approaches exist in the literature. Many of them~\cite{Donoho2005,Donoho2005a,Donoho2006b,Bruckstein2008TIT,Wang2009Allerton,Wang2011TSP,Khajehnejad2011,Zhao2014JORSC} are devoted to {\em modeling in the absence of noise}, focusing on questions such as the performance of convex optimization-based approaches for sparse recovery and the uniqueness of the sparse solution. Other works studied the theoretical performance of non-negative sparse modeling in the presence of noise~\cite{Wu2014,Slawski2011,Slawski2014}; however, those analyses focus on the specific case of either Gaussian or sub Gaussian noise.

In contrast to the statistical noise modeling used in previous analyses~\cite{Wu2014,Slawski2011,Slawski2014}, this paper considers the performance of non-negative sparse modeling under a more general scenario, where the observed signals have an unknown arbitrary distortion. Although our analysis can be applied to the additive random noise to obtain existing results, it is also immediately applicable to many other types of distortion, e.g., distortion present due to nonlinear mixing of the individual components. In the case of spectral unmixing, nonlinear mixing may come from nonlinear mixing of atoms~\cite{Dobigeon2014,Hapke2012} or spectral mismatches between the spectra of the minerals in the library and those involved in the observation~\cite{Fu2015c}.

Under this general scenario, we will investigate the conditions for successful reconstruction --- or regression --- of a signal using non-negative lasso by expanding previous analyses on the general lasso. Some of these studies consider dictionaries drawn according to a random distribution~\cite{Wainwright.ITIT2009a}. Others assume an arbitrary dictionary and pose conditions for successful regression that require a combinatorial amount of computation: examples include the spark, restricted isometry property, the restricted eigenvalue property, and the irrepresentable condition~\cite{Donoho.Elad.PNAS2003,CandesRIP,Zhao.Yu.JMLR2006,Bickel.etal.AS2009,cohen}. It is possible to relax the computational complexity of the verification process, using tools such as {\em coherence}~\cite{Donoho.Elad.PNAS2003,DonohoEladTemlyakov2006,Fuchs2005TIT,Tropp2004TIT,Tropp2006TIT}; however, all the aforementioned frameworks consider all possible combinations of atoms simultaneously, and therefore are found to often give very conservative assessments of regression performance. 

This paper follows the line of Tropp's work~\cite{Tropp2006TIT}. In contrast to the references above, Tropp has performed an analysis based on the so-called {\em exact recovery condition} (ERC), which provides conditions on successful reconstruction for all combinations of a fixed subset of atoms. The ERC can be easily computed and is compatible with well-known optimization-based and iterative greedy algorithms for sparse signal recovery and regression, and so it appears suitable for the analysis of non-negative versions. Furthermore, because the ERC is focused on sparse signals featuring a specific support (i.e., a set of indices for the signal's nonzero entries), its guarantees are less pessimistic than those provided by alternative approaches, which consider success for all sparse signals simultaneously, regardless of their support. Nonetheless, one could conceivably argue that restricting the set of signals of interest to a fixed support with non-negative entries may provide guarantees that are even closer to the actual performance of non-negative sparse recovery. While the proposed conditions require a specific set of atoms as an input, they are motivated by applications, such as hyperspectral unmixing, in which it is more useful to determine whether a specific set of atoms can be identified via sparsity-based methods, rather than provide guarantees on the recovery for all subsets, since many combinations of atoms might never materialize.

\subsection{Problem description}
Throughout this paper, we assume a linear model with non-negative coefficients,
\begin{IEEEeqnarray*}[]{c'l}\label{cvx:linmod}
	\bm{y} = \mat{A}\bm{x} + \bm{e} & (\bm{x}\succeq \bm{0})\IEEEyesnumber\label{eq:lmm}
\end{IEEEeqnarray*}
where $\bm{y} \in \mathbb{R}^L$ is an observation vector, $\mat{A} \in \mathbb{R}^{L\times N}$ is a dictionary matrix where the components $\bm{a}_j$ $(j=1,\ldots, N)$, called atoms, are stored in its columns, $\bm{x} \in \mathbb{R}^N$ is a non-negative coefficient vector, $\bm{e} \in \mathbb{R}^L$ is an error vector, and $\succeq$ (and its variants) denote element-wise inequality. Based on this model, we consider the problem of inferring atoms contributing the observation. In particular, we focus on the non-negative lasso (NLasso, also known as the non-negative Garrote in the statistics literature~\cite{Yuan2007}):
\begin{IEEEeqnarray}{l}\label{cvx:NLasso}
	\begin{IEEEeqnarraybox}[][c]{l'l}
		\underset{\bm{x}}{\text{minimize}}
		& \frac{1}{2}\|\bm{y}-\mat{A}\bm{x}\|_2^2 + \gamma \left\| \bm{x} \right\|_1 \\
		\text{subject to}
		& \bm{x} \succeq \bm{0},
	\end{IEEEeqnarraybox}
\end{IEEEeqnarray}
where $\gamma$ is a positive constant that controls the degree of sparsity. The weight $\gamma$ could be adaptively tuned for each element, e.g., as done in adaptive lasso~\cite{Zou2006}. NLasso has been used as a regression method in hyperspectral unmixing~\cite{Bioucas-Dias2010,Iordache2011TGRS}, as a variable selection method in economics~\cite{Wu2014}, and as a sparse recovery algorithm for face recognition~\cite{Ji2009ICMLA,He2013TNNLS,Cheng2013} and hyperspectral classification~\cite{Chen2011}. Many methods have been proposed for solving (\ref{cvx:NLasso}) such as non-negativity constrained least angle regression and selection~\cite{Morup2008}, full regularization path~\cite{Kim2013}, alternating direction algorithms~\cite{Bioucas-Dias2010}, iterative reweighted shrinkage~\cite{Iordache2011TGRS}, split Bregman~\cite{Szlam2010ICIP}, interior point~\cite{Kim2007}, and multiplicative updates~\cite{Wu2014}.

\subsection{Contribution of this paper}
We will derive model recovery conditions (MRCs) for NLasso (\ref{cvx:NLasso}). The MRCs allow us to predict if the correct atoms are identifiable via NLasso given a signal model for a specific set of atoms with noise or nonlinearity. The MRCs are reminiscent of ~\cite[Theorem 6]{Slawski2013} due the fact that both results are based on on the Karush-Kuhn-Tucker (KKT) conditions for convex optimization solutions. Our contribution are
\begin{enumerate}
	\item the development of MRCs in geometrically interpretable forms that are directly adopted to performance analysis of NLasso on any observation,
	\item the development of an approximately perfect MRC which not only guarantees correct signal recovery but also provides a ``practical converse'' that guarantees the failure of recovery in a practical sense.
\end{enumerate}%
Our MRCs indicate whether a certain distortion to a linear observation is tolerable by NLasso while succeeding in identifying the components of the dictionary being observed; for the specific case of nonlinear mixing, our result predicts accurately whether NLasso succeeds in component identification under nonlinear distortion, depending on its specific direction and magnitude. The approximately perfect MRC practically meets both necessity and sufficiency. Although the approximately perfect MRC is imperfect in a rigorous mathematical sense, it is quite powerful and in our experiments provides perfect prediction of the performance of NLasso. We also present some simplified variants of the approximately perfect MRC, which can be considered as customizations of Tropp's conditions~\cite{Tropp2006TIT} and are rigorously proved mathematically. Our criterion for perfect identification is that the atoms present in the observation are exactly identified by the algorithm (i.e., no missed atoms and no false alarms), without consideration for accuracy of the coefficient estimate values involved. 

We also showcase how our theorem can be used in real applications. More specifically, our theorem can predict whether NLasso will succeed in selecting the correct materials from a hyperspectral unmixing dictionary in the presence of deviations from the noiseless linear model, including measurement nonlinearities, bias, mismatch, and noise. Our experiments show that the approximately perfect MRC practically gives perfect assessment of the performance of NLasso.

\subsection{Mathematical notation}
\label{sec:mathnotation}
We specify the mathematical notation used in this paper. The support of $\bm{x} \in \mathbb{R}^N$ is the set of the indices associated with the non-zero elements of $\bm{x}$, denoted by $\supp{\bm{x}}$. $\mathcal{R}(\mat{X})$ is the range of the matrix $\mathbf{X}$. $\tr{\mat{M}}$, $\inv{\mat{M}}$, and $\mat{M}^\dagger$ denote the transpose, inverse, and
pseudoinverse of the matrix $\mat{M}$, respectively. $\|\mat{M}\|_{\infty, \infty}$ is an $(\infty, \infty)$ matrix operator norm and gives the maximum $\ell_1$-norm of the row vectors of $\mat{M}$. 

We denote a subset of the column indices of $\mat{A}\in \mathbb{R}^{L\times N}$ by $\Lambda \subseteq \{1,2,\ldots,N\}$, and the subdictionary that is composed of atoms associated with indices in $\Lambda$ by $\mat{A}_\Lambda$. Note that all of the theorems are discussed on a subset $\Lambda$ of the column indices as Tropp~\cite{Tropp2006TIT} did. For any coefficient vector $\bm{x} \in \mathbb{R}^N$ defined in~\eqref{eq:lmm}, we denote the vector composed of the elements of $\bm{x}$ indexed by $\Lambda$ by $\bm{x}_\Lambda$. We also denote the whole column index set $\Omega = \{1,2,\ldots,N\}$, and the complement of $\Lambda$ by $\stcomp{\Lambda} = \Omega\!\setminus\!\Lambda$ where $\setminus$ is the difference of two sets. 

\subsection{Organization of this paper}
The rest of this paper is organized as follows. Section~\ref{sec:previouswork} introduces a sufficient MRC shown in our previous work~\cite{Yuki2015Whispers}. Section~\ref{sec:NLasso} demonstrates several MRCs for NLasso of varying tightness. Section~\ref{sec:application} illustrates an application of our MRCs to a hyperspectral unmixing task and Section~\ref{sec:conclusion} concludes this paper.

\section{Previous work}
\label{sec:previouswork}
We start our discussion with our previous work~\cite{Yuki2015Whispers}, which introduced a sufficient MRC to guarantee correct model recovery using NLasso. 
The previous work stands on the work by Tropp~\cite{Tropp2006TIT} and depends on the exact recovery coefficient (ERC), defined by
\begin{equation}
\mathrm{ERC}(\Lambda) := 1-\max_{n \notin \Lambda}\|\mat{A}_\Lambda^\dagger \bm{a}_n\|_1\label{eq:erc}.
\end{equation}
Note that it is implicitly assumed that the columns of $\mat{A}_\Lambda$ are linearly independent so that the pseudoinverse exists. Broadly speaking, the coefficient evaluates how far the atoms outside of $\Lambda$ are from the convex hull determined by the atoms in $\Lambda$ and their antipodes. Intuitively, a larger $\mathrm{ERC}$ is preferred because it reduces correlation between $\mat{A}_{\Lambda}$ and atoms outside the set. The following theorem provides performance guarantees for the lasso that are specific to a particular support $\Lambda$. Tropp considers the general lasso
\begin{IEEEeqnarray*}{l}
	\begin{IEEEeqnarraybox}[][c]{l'l}
		\underset{\bm{x}}{\text{minimize}}
		& \frac{1}{2}\|\bm{y}-\mat{A}\bm{x}\|_2^2 + \gamma \left\| \bm{x} \right\|_1 \\
	\end{IEEEeqnarraybox}\IEEEyesnumber\label{cvx:PL-lasso}
\end{IEEEeqnarray*}
and derived a theorem to guarantee the performance of this problem:
\vspace{5pt}
\begin{theorem}\cite[Theorem 8]{Tropp2006TIT}
	Let $\Lambda$ index a linearly independent collection of columns of $\mathbf{A}$ for which $\mathrm{ERC}(\Lambda) \ge 0$. Suppose that $\bm{y}$ is an input signal whose $\ell_2$ best approximation $\bm{y}_\Lambda = \mathbf{A}_\Lambda \mathbf{A}_\Lambda^\dagger \bm{y}$ over $\mathbf{A}_\Lambda$ satisfies the correlation condition
	\begin{align}
	\|\tr{\mat{A}}(\bm{y}-\bm{y}_\Lambda)\|_\infty \le \gamma \mathrm{ERC}(\Lambda).
	\label{eq:bpdncorcond}
	\end{align}
	Let $\bm{x}^\star$ be the solution of the lasso~\eqref{cvx:PL-lasso} with parameter $\gamma$. We may conclude the following.
	\begin{itemize}
		\item $\mathrm{supp}(\bm{x}^\star)$, is contained in $\Lambda$;
		\item the distance between $\bm{x}^\star$ and the optimal coefficient vector $\bm{c}_\Lambda = \mathbf{A}_\Lambda^\dagger \bm{y}$ (appropriately zero-padded) satisfies
		\begin{align}
		\|\bm{x}^\star_\Lambda-\bm{c}_\Lambda\|_\infty \le \gamma \|\inv{(\tr{\mat{A}}_\Lambda\mat{A}_\Lambda)}\|_{\infty,\infty};
		\end{align}
		\item and $\mathrm{supp}(\bm{x}^\star)$ contains the indices $\lambda \in \Lambda$ for which
		\begin{align}
		|\bm{c}_\Lambda(\lambda)| > \gamma\|\inv{(\tr{\mat{A}}_\Lambda\mat{A}_\Lambda)}\|_{\infty,\infty}.
		\end{align}
	\end{itemize}
	\label{thm:Tropp_MRC_lasso}
\end{theorem}
\vspace{5pt}
Using Theorem~\ref{thm:Tropp_MRC_lasso}, we can derive the following theorem to guarantee the performance of NLasso:
\begin{theorem}\cite[Theorem 2]{Yuki2015Whispers}\label{thm:SMRC_NLasso_ERC-based}
	Assume a signal model $\bm{y} = \mat{A}\bm{x}^{\text{true}}+\bm{e}$, where the 
	abundance vector $\bm{x}^{\text{true}} \succeq \bm{0}$, $\Lambda = 
	\mathrm{supp}(\bm{x}^{\text{true}})$ indexes a linearly independent collection of 
	columns of $\mat{A}$, and $\bm{e}$ represents the effect of noise or 
	nonlinear distortion during acquisition. Let $\opt{\bm{x}}$ be the solution 
	of NLasso with parameter $\gamma$. If the noise $\bm{e}$ obeys
	\begin{IEEEeqnarray*}{c}
		\|\tr{\mat{A}}{\mat{P}_\Lambda^\perp}\bm{e}\|_\infty \le 
		\gamma \mathrm{ERC}(\Lambda)\IEEEyesnumber\label{eq:SMRC_NLasso_ERC-based:NSCC}
	\end{IEEEeqnarray*}
	where $\mat{P}_{\Lambda}^\perp$ is the projector onto the orthogonal complement of $\mathcal{R}(\mat{A}_{\Lambda})$, and 
	\begin{IEEEeqnarray*}{c}
		\bm{x}_{\Lambda}^{\text{true}} \succ \gamma\|\inv{(\tr{\mat{A}}_\Lambda\mat{A}_{\Lambda})}\|_{\infty,\infty} -\mat{A}_{\Lambda}^\dagger \bm{e}, \IEEEyesnumber
		\label{eq:SMRC_NLasso_ERC-based:MCC}
	\end{IEEEeqnarray*}
	then we have that $\mathrm{supp}(\opt{\bm{x}}) = \Lambda$.
\end{theorem}
\begin{IEEEproof}
	We begin by considering the solution $\bm{x}^\star$ to the lasso with parameter $\gamma$ for the input $\bm{y}$. By applying Theorem~\ref{thm:Tropp_MRC_lasso} and seeing that
	\ifCLASSOPTIONdraftcls
		\begin{IEEEeqnarray}{r,c,l}
			\|\tr{\mat{A}}(\bm{y}-\bm{y}_\Lambda)\|_\infty
			&=& \|\tr{\mat{A}}(\bm{y}-\mat{A}_\Lambda{\mat{A}_\Lambda^\dagger}\bm{y})\|_\infty \IEEEnonumber\\
			&=& \|\tr{\mat{A}}(\mat{A}\bm{x}^\text{true}+\bm{e}-\mat{A}_\Lambda{\mat{A}_\Lambda^\dagger}(\mat{A}\bm{x}^{\text{true}}+\bm{e}))\|_\infty\IEEEnonumber\\
			&=& \|\tr{\mat{A}}(\mat{A}_\Lambda\bm{x}_\Lambda^{\text{true}}+\bm{e}-\mat{A}_\Lambda{\mat{A}_\Lambda^\dagger}(\mat{A}_\Lambda\bm{x}_\Lambda^{\text{true}}+\bm{e}))\|_\infty\IEEEnonumber\\
			&=& \|\tr{\mat{A}}(\mat{A}_\Lambda\bm{x}_\Lambda^{\text{true}}+\bm{e}-\mat{A}_\Lambda\bm{x}_\Lambda^{\text{true}}-\mat{A}_\Lambda \mat{A}_\Lambda^\dagger \bm{e})\|_\infty \IEEEnonumber\\ 
			&=& \|\tr{\mat{A}}(\bm{e}-\mat{A}_\Lambda \mat{A}_\Lambda^\dagger \bm{e})\|_\infty =  \|\tr{\mat{A}}(\mat{I}-\mat{A}_\Lambda \mat{A}_\Lambda^\dagger) \bm{e}\|_\infty\IEEEnonumber\\
			&=& \|\tr{\mat{A}}\mat{P}_{\Lambda}^\perp\bm{e}\|_\infty, \label{eq:proofnoise}
		\end{IEEEeqnarray}
	\else
		\begin{IEEEeqnarray}{r,c,l}
			&&\|\tr{\mat{A}}(\bm{y}-\bm{y}_\Lambda)\|_\infty \IEEEnonumber\\
			&=& \|\tr{\mat{A}}(\bm{y}-\mat{A}_\Lambda{\mat{A}_\Lambda^\dagger}\bm{y})\|_\infty \IEEEnonumber\\
			&=& \|\tr{\mat{A}}(\mat{A}\bm{x}^\text{true}+\bm{e}-\mat{A}_\Lambda{\mat{A}_\Lambda^\dagger}(\mat{A}\bm{x}^{\text{true}}+\bm{e}))\|_\infty\IEEEnonumber\\
			&=& \|\tr{\mat{A}}(\mat{A}_\Lambda\bm{x}_\Lambda^{\text{true}}+\bm{e}-\mat{A}_\Lambda{\mat{A}_\Lambda^\dagger}(\mat{A}_\Lambda\bm{x}_\Lambda^{\text{true}}+\bm{e}))\|_\infty\IEEEnonumber\\
			&=& \|\tr{\mat{A}}(\mat{A}_\Lambda\bm{x}_\Lambda^{\text{true}}+\bm{e}-\mat{A}_\Lambda\bm{x}_\Lambda^{\text{true}}-\mat{A}_\Lambda \mat{A}_\Lambda^\dagger \bm{e})\|_\infty \IEEEnonumber\\ 
			&=& \|\tr{\mat{A}}(\bm{e}-\mat{A}_\Lambda \mat{A}_\Lambda^\dagger \bm{e})\|_\infty =  \|\tr{\mat{A}}(\mat{I}-\mat{A}_\Lambda \mat{A}_\Lambda^\dagger) \bm{e}\|_\infty\IEEEnonumber\\
			&=& \|\tr{\mat{A}}\mat{P}_{\Lambda}^\perp\bm{e}\|_\infty, \label{eq:proofnoise}
		\end{IEEEeqnarray}
	\fi
	we have that \eqref{eq:proofnoise} and \eqref{eq:SMRC_NLasso_ERC-based:NSCC} imply \eqref{eq:bpdncorcond}. Thus, we obtain the following results:
	\begin{itemize}
		\item The support of $\bm{x}^\star$, $\mathrm{supp}(\bm{x}^\star)$, is contained in $\Lambda$, and
		\item the distance between $\bm{x}^\star$ and the optimal coefficient vector 
		\begin{IEEEeqnarray}{r,c,l}
			\bm{c}_\Lambda &=& \mat{A}_\Lambda^\dagger \bm{y} = \mat{A}_\Lambda^\dagger (\mat{A}\bm{x}^{\text{true}}+\bm{e}) = \mat{A}_\Lambda^\dagger (\mat{A}_\Lambda \bm{x}_\Lambda^{\text{true}}+\bm{e})\IEEEnonumber*\\
			&=& \bm{x}_\Lambda^\text{true}+\mat{A}_\Lambda^\dagger \bm{e}
		\end{IEEEeqnarray}
		(appropriately zero-padded) satisfies
		\begin{align}
		\|\bm{x}^\star-\bm{x}_\Lambda^{\text{true}}-\mat{A}_\Lambda^\dagger \bm{e}\|_\infty \le \gamma \|\inv{(\mat{A}_\Lambda^T\mat{A}_\Lambda)}\|_{\infty,\infty}.
		\label{eq:inftynoise}
		\end{align}
	\end{itemize}
	The result \eqref{eq:inftynoise} implies that for each $n \in \Lambda$ we have
	\ifCLASSOPTIONdraftcls
	\begin{IEEEeqnarray}{r,c,l}
		|\bm{x}^\star(n)-(\bm{x}^{\text{true}}(n)+\bm{w}(n))| &\le& \gamma \|\inv{(\tr{\mat{A}}_\Lambda\mat{A}_\Lambda)}\|_{\infty,\infty},\IEEEnonumber*\\
		-\gamma \|\inv{(\tr{\mat{A}}_\Lambda\mat{A}_\Lambda)}\|_{\infty,\infty} &\le& \bm{x}^\star(n)-\bm{x}^{\text{true}}(n)-\bm{w}(n),\\ 
		\bm{x}^{\text{true}}(n)+\bm{w}(n)-\gamma \|\inv{(\tr{\mat{A}}_\Lambda\mat{A}_\Lambda)}\|_{\infty,\infty} &\le& \bm{x}^\star(n) ,
	\end{IEEEeqnarray}
	\else
	\begin{IEEEeqnarray}{c}
		|\bm{x}^\star(n)-(\bm{x}^{\text{true}}(n)+\bm{w}(n))| \le \gamma \|\inv{(\tr{\mat{A}}_\Lambda\mat{A}_\Lambda)}\|_{\infty,\infty},\IEEEnonumber*\\
		-\gamma \|\inv{(\tr{\mat{A}}_\Lambda\mat{A}_\Lambda)}\|_{\infty,\infty} \le \bm{x}^\star(n)-\bm{x}^{\text{true}}(n)-\bm{w}(n),\\ 
		\bm{x}^{\text{true}}(n)+\bm{w}(n)-\gamma \|\inv{(\tr{\mat{A}}_\Lambda\mat{A}_\Lambda)}\|_{\infty,\infty}\le\bm{x}^\star(n) ,
	\end{IEEEeqnarray}
	\fi
	where we denote $\bm{w} = \mathbf{A}_\Lambda^\dagger \bm{e}$.
	Thus, from the condition \eqref{eq:SMRC_NLasso_ERC-based:MCC}, we have that $\bm{x}^\star(n) > 0$ for all $n \in \Lambda$, which implies that $\Lambda \subseteq \mathrm{supp}(\bm{x}^\star)$. 
	Furthermore, since $\mathrm{supp}(\bm{x}^\star) \subseteq \Lambda$, then we have that $\mathrm{supp}(\bm{x}^\star) = \Lambda$ and so it follows that $\bm{x}^\star \succ 0$, i.e., the solution of the lasso is non-negative. 
	This implies that the solution of NLasso for the same input and parameter value obeys $\opt{\bm{x}} = \bm{x}^\star$ (i.e., the solution of NLasso matches the solution of the unconstrained lasso), and so $\mathrm{supp}(\opt{\bm{x}}) = \mathrm{supp}(\bm{x}^\star) = \Lambda =  \mathrm{supp}(\bm{x}^{\text{true}})$.
\end{IEEEproof}

The sufficient condition is composed of two inequalities; the first one~\eqref{eq:SMRC_NLasso_ERC-based:NSCC} explains how much deviation from linearity is allowed, and the second one~\eqref{eq:SMRC_NLasso_ERC-based:MCC} shows the minimum value of the coefficient to be detected.
This condition is a demanding sufficient MRC, as shown in~\cite{Yuki2015Whispers}. More specifically, there are still many observations on which NLasso is successful, but for which the condition is not met.

\section{Main results}
\label{sec:NLasso}
\subsection{Fundamental results for NLasso}
\label{sec:fund_nlasso}
We note that NLasso is equivalent to
\begin{IEEEeqnarray}{l}\label{cvx:NLasso2}
	\begin{IEEEeqnarraybox}[][c]{l'l}
		\underset{\bm{x}}{\text{minimize}}
		& \frac{1}{2}\|\bm{y}-\mat{A}\bm{x}\|_2^2 + \gamma \tr{\bm{1}}_N\bm{x} \\
		\text{subject to}
		& \bm{x} \succeq \bm{0},
	\end{IEEEeqnarraybox}
\end{IEEEeqnarray}
where $\bm{1}_N$ is the $N$ length vector with all elements being one. This minimization problem becomes NNLS when $\gamma=0$. First, we provide MRCs for which the subset of atoms $\Lambda$ contains the support of minimizers of NLasso. In particular, we will give a condition for which a solution to the restricted NLasso
\begin{IEEEeqnarray*}{l}
	\begin{IEEEeqnarraybox}[][t]{l'l}
		\underset{\bm{v}_\Lambda}{\text{minimize}}
		& \frac{1}{2}\|\bm{y}-\mat{A}_\Lambda\bm{v}_\Lambda\|_2^2 + \gamma \tr{\bm{1}}_J\bm{v}_\Lambda \\
		\text{subject to}
		& \bm{v}_\Lambda \succeq \bm{0},
	\end{IEEEeqnarraybox}
	\IEEEyesnumber\label{cvx:NLasso_Lambda}
\end{IEEEeqnarray*}
also becomes a solution to the original NLasso~\eqref{cvx:NLasso}. The condition is given by the following theorem. 
\vspace{5pt}
\begin{theorem}\label{thm:APMRC_NLasso_base}
	Let $\Lambda$ be a subset of column indices of the dictionary matrix $\mat{A}$ such that $|\Lambda| = J \le N$. 
	If the inequalities
	\begin{IEEEeqnarray*}[]{c'l}
		\tr{(\bm{y}-\mat{A}_\Lambda\opt{\bm{v}}_\Lambda)}\bm{a}_j < \gamma &
		\text{for all }j \in \stcomp{\Lambda}\IEEEyesnumber\label{eq:APMRC_NLasso_base}
	\end{IEEEeqnarray*}
	hold for all solutions $\opt{\bm{v}}_\Lambda \in \mathbb{R}^{J}$ of the restricted NLasso~\eqref{cvx:NLasso_Lambda} over the column subset $\Lambda$, then all solutions to the general NLasso~\eqref{cvx:NLasso} have their supports contained in $\Lambda$. 
\end{theorem}
\vspace{5pt}
A proof of this theorem is found in the appendix. This theorem states a condition for which the restricted NLasso~\eqref{cvx:NLasso_Lambda} yields a global solution of the original problem~\eqref{cvx:NLasso}. 
Although this condition requires knowledge of the solutions of the restricted NLasso, the theorem considers quite general cases:
\begin{enumerate}
	\item the subdictionary $\mat{A}_\Lambda$ can have linearly dependent columns,
	\item the restricted NLasso over columns in $\Lambda$ can have multiple minimizers,
	\item the columns of $\mat{A}$ are not restricted to be normalized.
\end{enumerate}
Thus, the theorem serves as a fundamental result to derive other practical MRCs in subsequent sections. When atoms associated with indices in $\Lambda$ are linearly independent to each other, we can further assume the uniqueness of the solution because the restricted problem has the unique solution.

The condition~\eqref{eq:APMRC_NLasso_base} is a sufficient but not necessary condition for the event $\mathrm{supp}(\opt{\bm{x}})\subseteq \Lambda$. However, \eqref{eq:APMRC_NLasso_base} is quite close to a necessary condition, as shown in the following theorem.
\vspace{5pt}
\begin{theorem}\label{thm:APMRC_NLasso_base_eq}
	Under the assumption of Theorem~\ref{thm:APMRC_NLasso_base}, if the support $\mathrm{supp}(\opt{\bm{x}})$ of each solution $\opt{\bm{x}}$ 
	to the general NLasso~\eqref{cvx:NLasso} is contained in $\Lambda$, then the following condition holds for all solutions $\opt{\bm{v}}_\Lambda \in \mathbb{R}^{J}$ of the restricted NLasso~\eqref{cvx:NLasso_Lambda} over the column subset $\Lambda$:
	\begin{IEEEeqnarray*}[]{c'l}
		\tr{(\bm{y}-\mat{A}_\Lambda\opt{\bm{v}}_\Lambda)}\bm{a}_j \le \gamma &
		\text{for all }j \in \stcomp{\Lambda}.\IEEEyesnumber\label{eq:APMRC_NLasso_base_eq}
	\end{IEEEeqnarray*}
\end{theorem}
\vspace{5pt}
The proof of this theorem is found in the appendix. This theorem indicates that the condition~\eqref{eq:APMRC_NLasso_base_eq} is a necessary condition for $\mathrm{supp}(\opt{\bm{x}})\subseteq \Lambda$. Hence, a necessary and sufficient condition for the event $\mathrm{supp}(\opt{\bm{x}})\subseteq \Lambda$ lies somewhere between \eqref{eq:APMRC_NLasso_base} and \eqref{eq:APMRC_NLasso_base_eq}. More specifically, equalities need to be added to \eqref{eq:APMRC_NLasso_base} only for some indices $j$ to obtain a necessary and sufficient condition. Nonetheless, it is worth noting that the cases in which \eqref{eq:APMRC_NLasso_base_eq} holds with equality will be rare in practice. Therefore, the conditions \eqref{eq:APMRC_NLasso_base} and \eqref{eq:APMRC_NLasso_base_eq} are practically identical, implying that \eqref{eq:APMRC_NLasso_base} is a practically valid necessary and sufficient condition for the event $\mathrm{supp}(\opt{\bm{x}})\subseteq \Lambda$.

\begin{remark}
Just like NLasso becomes NNLS when $\gamma=0$, a restricted NNLS problem is given by the minimization problem~\eqref{cvx:NLasso_Lambda} when $\gamma=0$. As a special case of Theorem~\ref{thm:APMRC_NLasso_base}, we can define an optimal condition for NNLS specific to an index set $\Lambda$:
\vspace{5pt}
\begin{corollary}\label{thm:APMRC_NNLS_base}
	Let $\Lambda$ be a subset of column indices of the dictionary matrix $\mat{A}$ such that $|\Lambda| = J \le N$. 
	If the inequalities
	\begin{IEEEeqnarray*}[]{c'l}
		\tr{(\bm{y}-\mat{A}_\Lambda\opt{\bm{v}}_\Lambda)}\bm{a}_j < 0 &
		\text{for all }j \in \stcomp{\Lambda}\IEEEyesnumber\label{eq:APMRC_NNLS_base}
	\end{IEEEeqnarray*}
	hold for the solution $\opt{\bm{v}}_\Lambda \in \mathbb{R}^{J}$ of the restricted NNLS problem over the column subset $\Lambda$, then all solutions to the general NNLS have their supports contained in $\Lambda$. 
\end{corollary}
\vspace{5pt}
We note in passing that the set of inequalities \eqref{eq:APMRC_NNLS_base} is identical to one of the stopping criteria introduced for non-negative orthogonal matching pursuit~\cite{Bruckstein2008,Yaghoobi2015}, an alternative algorithm for non-negative sparse signal recovery. 
\end{remark}

\subsection{Approximately Perfect MRC for NLasso}
\label{sec:APMRC_NLasso}
\begin{figure}[!t]
	\centering
	\includegraphics[width=200pt]{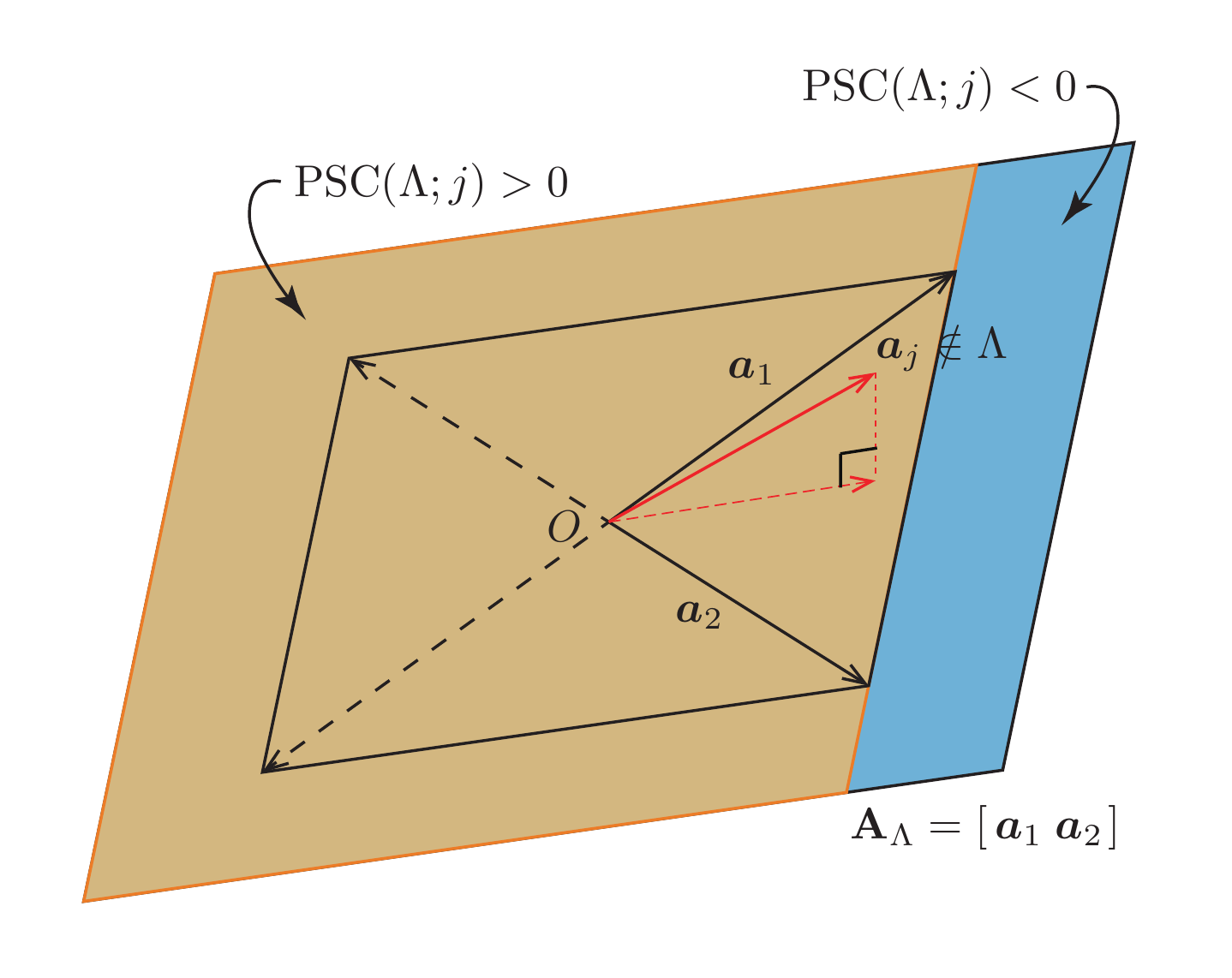}%
	\caption{Illustration of geometric interpretation of $\mathrm{PSC}$.}\label{fig:psc}
\end{figure}
\begin{figure}[!t]
	\centering
	\includegraphics[width=200pt]{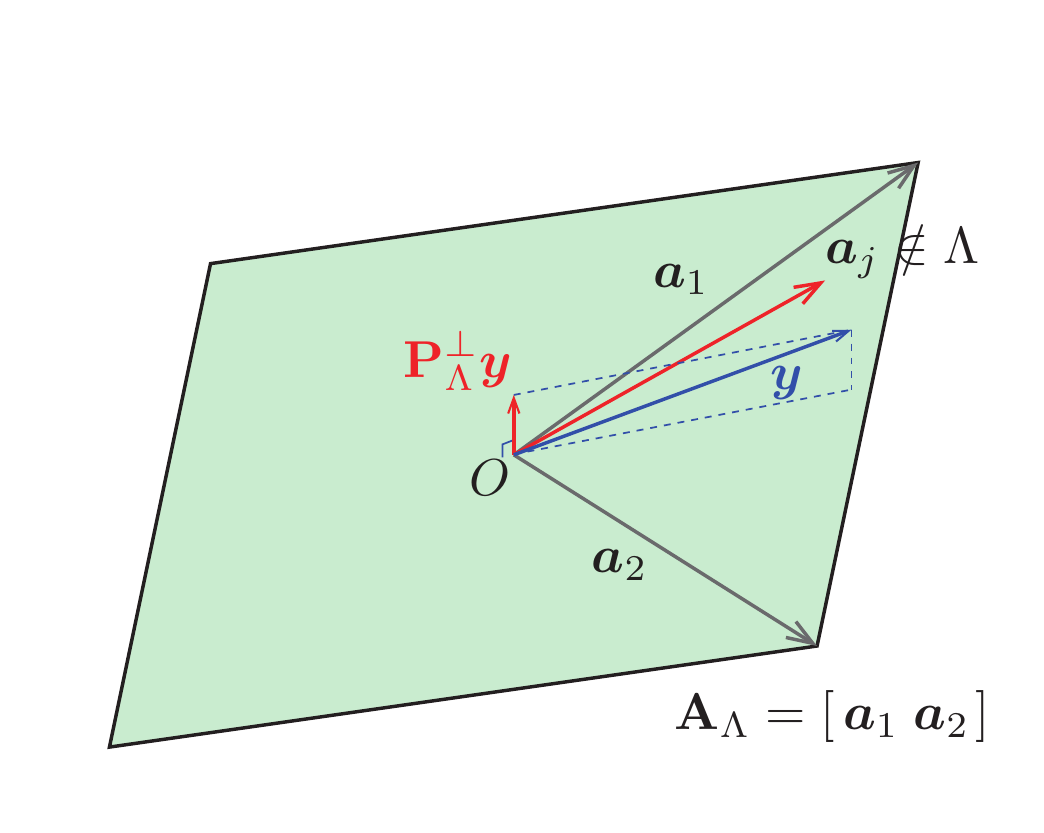}%
	\caption{Illustration of geometric interpretation of NSCC.\label{fig:nscc}}
\end{figure}

This section provides MRCs for which the subset of atoms $\Lambda$ exactly matches the support of the minimizers of NLasso. We again assume that the atoms associated with indices in $\Lambda$ are linearly independent. First, we define a metric that we call {\it positive subset coherence} (PSC):
\begin{IEEEeqnarray}[]{c}
	\mathrm{PSC}(\Lambda;j) := 1-\tr{\bm{1}_{J}}{\mat{A}}_{\Lambda}^{\dagger} \bm{a}_j.
\end{IEEEeqnarray}
The $\mathrm{PSC}$ measures how positively aligned the $j^\textrm{th}$ atom $\bm{a}_j$ in the library is to the convex cone determined by the columns of $\mat{A}_\Lambda$. The index $j$ is usually selected from outside $\Lambda$. Figure~\ref{fig:psc} illustrates a geometric interpretation of $\mathrm{PSC}$ focusing on when the sign of $\mathrm{PSC}$ changes. The $\mathrm{PSC}$ becomes positive when the orthogonal projection of $\bm{a}_j$ onto the subspace spanned by the column vectors of $\mat{A}_\Lambda$ falls on the same side of the hyperplane passing through the column vectors of $\mat{A}_\Lambda$ as the origin; negative when the origin and the column $\bm{a}_j$ are on opposite sides of the aforementioned hyperplane; and zero when $\bm{a}_j$ is contained in this hyperplane. This $\mathrm{PSC}$ plays an important role in the next approximately perfect MRC.
\vspace{3pt}
\begin{theorem}[Approximately Perfect MRC for NLasso]\label{thm:APMRC_NLasso}
	Let $\Lambda$ be a subset of the column indices of the dictionary matrix $\mat{A}$ such that $|\Lambda| = J \le \min(L,N)$ and the atoms associated with indices in $\Lambda$ are linearly independent. Let $\opt{\bm{x}}$ be a solution to NLasso. The support of $\opt{\bm{x}}$, $\mathrm{supp}(\opt{\bm{x}})$, is equal to $\Lambda$ if the following two conditions hold:
	\vspace{-3pt}
	\ifCLASSOPTIONdraftcls
		\begin{IEEEeqnarray*}[]{'L,l}
			1) & \text{Minimum coefficient condition (MCC):} \\
			& \hspace{40pt}\mat{A}_{\Lambda}^\dagger\bm{y} \succ \gamma\inv{\bigl(\tr{\mat{A}}_{\Lambda}\mat{A}_{\Lambda}\bigr)}\bm{1}_J\IEEEyesnumber\label{eq:APMRC_NLasso:MCC}\\
			2) & \text{Non-linearity vs. Subset Coherence Condition (NSCC):} \\
			& \hspace{40pt}\tr{\bm{y}}\mat{P}^{\perp}_{\Lambda}\bm{a}_j < \gamma \mathrm{PSC}(\Lambda;j)  \quad \text{ for all } j \in \stcomp{\Lambda}.\IEEEyesnumber\label{eq:APMRC_NLasso:NSCC}
		\end{IEEEeqnarray*}
	\else
		\begin{IEEEeqnarray*}[]{-L,l}
			1) & \text{Minimum coefficient condition (MCC):} \\
			& \hspace{40pt}\mat{A}_{\Lambda}^\dagger\bm{y} \succ \gamma\inv{\bigl(\tr{\mat{A}}_{\Lambda}\mat{A}_{\Lambda}\bigr)}\bm{1}_J\IEEEyesnumber\label{eq:APMRC_NLasso:MCC}\\
			2) & \text{Non-linearity vs. Subset Coherence Condition (NSCC):} \\
			& \hspace{40pt}\tr{\bm{y}}\mat{P}^{\perp}_{\Lambda}\bm{a}_j < \gamma \mathrm{PSC}(\Lambda;j)  \quad \text{ for all } j \in \stcomp{\Lambda}.\IEEEyesnumber\label{eq:APMRC_NLasso:NSCC} \vspace{-2pt}
		\end{IEEEeqnarray*}
	\fi
	Furthermore, the minimizer $\opt{\bm{x}}$ is equal to the appropriate zero-padding of the solution to the restricted NLasso\vspace{-2pt}
	\begin{IEEEeqnarray*}[]{c}
		\opt{\bm{v}}_{\Lambda} = \mat{A}_{\Lambda}^\dagger\bm{y} - \gamma\inv{\bigl(\tr{\mat{A}}_{\Lambda}\mat{A}_{\Lambda}\bigr)}\bm{1}_J.
	\end{IEEEeqnarray*}
\end{theorem}
The proof of this theorem is found in~Section~\ref{sec:proof:PMRC_NLasso_specific}. The MCC measures whether the entries of the least squares solution $(\mat{A}_\Lambda^\dagger\bm{y})$ are sufficiently large. The NSCC specifies the degree of nonlinear distortion that can be tolerated with respect to each $\mathrm{PSC}$ and dictionary atom $\bm{a}_j$. The left hand side of~\eqref{eq:APMRC_NLasso:NSCC} is the inner product between $\bm{a}_j$ and the orthogonal projection of $\bm{y}$ onto the orthogonal complement of $\mathcal{R}(\mat{A}_\Lambda)$. The latter projection can be interpreted as nonlinear noise because it is considered as the deviation of the observation $\bm{y}$ from the span of $\mat{A}_\Lambda$. Figure~\ref{fig:nscc} shows a geometric interpretation of the NSCC. As explained, the right hand side of~\eqref{eq:APMRC_NLasso:NSCC} quantifies the alignment of $\bm{a}_j$ with the convex cone obtained from $\mat{A}_\Lambda$. Because $\gamma$ is usually positive, a larger $\mathrm{PSC}$ relaxes the upper bound of $\tr{\bm{y}}\mat{P}^{\perp}_{\Lambda}\bm{a}_j$. Thus, it is preferable for $\bm{a}_j$ to be less aligned to the columns of $\mat{A}_\Lambda$.

The inequality~\eqref{eq:APMRC_NLasso:MCC} needs to strictly hold for mathematical rigor because of the definition of the support (the set of indices that have non-zero entries). If we instead allowed for equality in~\eqref{eq:APMRC_NLasso:MCC}, we would not be able to guarantee that $\Lambda$ exactly matches the support of $\opt{\bm{x}}$ since some of the entries in $\opt{\bm{v}}_{\Lambda}$ might be zero-valued, cf. eq.~(\ref{eq:sol_NLasso_specific}), although the solution $\opt{\bm{x}}$ would be still expressed in the same way, i.e., as the appropriate zero-padding of the solution $\opt{\bm{v}}_{\Lambda}$ to the restricted NLasso. Nonetheless, equality can be added in practice since the event for which the equations hold with equality is quite rare, as described in the discussion of Section~\ref{sec:fund_nlasso}. This discussion is also true for the inequality~\eqref{eq:APMRC_NLasso:NSCC}; equality could be added to the inequality in practical settings.

Interestingly, Conditions \eqref{eq:SMRC_NLasso_ERC-based:NSCC} and \eqref{eq:SMRC_NLasso_ERC-based:MCC} in Theorem~\ref{thm:SMRC_NLasso_ERC-based} are similar to the NSCC and MCC, but the former are not specific to particular indices $j$. This structure of the condition is shared with the simplified sufficient conditions derived in Section~\ref{sec:SMRCs}.

\begin{remark}
	Theorem~\ref{thm:APMRC_NLasso} can be specialized to specific noise models. In the case of random noise (e.g., Gaussian or subgaussian), it is possible to obtain the likelihood of the NSCC being met in terms of the noise variance; such a result matches~\cite[Theorem 6]{Slawski2013}. Under a linear noise model (e.g., the distortion corresponds to a linear combination of the atoms in lies in $\mat{A}_\Lambda$), the left hand side of \eqref{eq:APMRC_NLasso:NSCC} always becomes equal to zero, and it suffices to require a non-negative $\mathrm{PSC}$ for each of the atoms indexed in $\stcomp{\Lambda}$. In all other cases, the NSCC allows us to distinguish between tolerable and intolerable distortions. If the nonlinear distortion (i.e., the portion of the distortion in the space orthogonal to $\mathcal{R}(\mat{A}_\Lambda)$) forms obtuse angles with the atoms indexed by $\stcomp{\Lambda}$, the NSCC will be satisfied regardless of the magnitude of the nonlinearity. Conversely, for atoms for which these angles are acute, the value of the $\mathrm{PSC}$ for the specific atom will dictate the tolerable magnitude of the projection of the nonlinear distortion onto the atom. 
\end{remark}

\begin{remark}
We note that meeting the MCC and NSCC will also depend on the value of the trade-off parameter $\gamma$. Intuitively, the chance for false alarm increases as $\gamma$ decreases (promoting denser solutions) and missed detection is more likely to occur as $\gamma$ increases (promoting sparser solutions). The MCC provides an upper bound on $\gamma$ needed to avoid missed detections, while the NSCC provides a lower bound needed to prevent false alarms. The bounds will depend on the observation $\bm{y}$ and support $\Lambda$, which indicates that the performance of NLasso can be improved by adaptively optimizing $\gamma$. It is also easy to see that one can formulate configurations $(\bm{y},\Lambda)$ for which no value of $\gamma$ meets both the MCC and NSCC.
\end{remark}

\begin{remark}
Theorem~\ref{thm:APMRC_NLasso} can also be specialized to NNLS (e.g., $\gamma = 0$), providing the following corollary. 
\begin{corollary}[Approximately Perfect MRC for NNLS]\label{thm:APMRC_NNLS}
	Let $\Lambda$ be a subset of column indices of the dictionary matrix $\mat{A}$ such that $|\Lambda| = J \le \min(L,N)$ and the atoms associated with indices in $\Lambda$ are linearly independent. Let $\opt{\bm{x}}$ be the solution of NNLS. The support of $\opt{\bm{x}}$, $\mathrm{supp}(\opt{\bm{x}})$, is equal to $\Lambda$ if $\mat{A}_{\Lambda}^\dagger\bm{y} \succ 0$ and $\max_{j}{\tr{\bm{y}}\mat{P}^{\perp}_{\Lambda}\bm{a}_j} < 0$. 
\end{corollary}
In this case, the MCC requires for the restricted least squares solution to be non-negative, while the NSCC requires all angles between the atoms indexed in $\stcomp{\Lambda}$ and the nonlinear distortion in the orthogonal space of $\mathcal{R}(\mat{A}_\Lambda)$ to be obtuse. 
\end{remark}

\subsection{Proof of Theorem~\ref{thm:APMRC_NLasso}}
\label{sec:proof:PMRC_NLasso_specific}

Recall that the atoms associated with the index set $\Lambda$ are linearly independent to each other and the two conditions~\eqref{eq:APMRC_NLasso:MCC} and \eqref{eq:APMRC_NLasso:NSCC} hold. First let us consider the restricted NLasso defined by~\eqref{cvx:NLasso_Lambda}. The problem~\eqref{cvx:NLasso_Lambda} has the unique minimizer because the objective function is strictly convex and the domain is a convex region. The Lagrangian of~\eqref{cvx:NLasso_Lambda} is given by
\begin{IEEEeqnarray*}[]{c}
	L(\bm{v}_\Lambda, \bm{\lambda}) = \frac{1}{2}\|\bm{y}-\mat{A}_\Lambda\bm{v}_\Lambda\|_2^2 + \gamma \tr{\bm{1}_J}\bm{v}_\Lambda - \tr{\bm{\lambda}}\bm{v}_\Lambda,
\end{IEEEeqnarray*}
where $\bm{\lambda} \in \mathbb{R}^{J}$ is a Lagrange multiplier with $\bm{\lambda} \succeq \bm{0}$. According to~\cite[Theorem 28.3, p.~281]{ConvexAnalysis}, $\opt{\bm{v}}_\Lambda$ and $\opt{\bm{\lambda}}$ are optimal if and only if they satisfy the KKT conditions
\begin{IEEEeqnarray*}[]{-s'L}
	\;\;1) & \opt{\bm{v}}_\Lambda,\opt{\bm{\lambda}} \succeq \bm{0} \text{ and } \opt{\bm{\lambda}}(n)\opt{\bm{v}}_\Lambda(n) = 0 \text{ for all } n=1,\ldots,J \\
	\;\;2) & \bm{0} = \partial L(\bm{v}_\Lambda,\opt{\bm{\lambda}}) / \partial \bm{v}_\Lambda|_{\bm{v}_\Lambda=\opt{\bm{v}}_\Lambda}.
\end{IEEEeqnarray*}
Those conditions are true for $\opt{\bm{v}}_{\Lambda} = \mat{A}_{\Lambda}^\dagger\bm{y} - \gamma\inv{\bigl(\tr{\mat{A}}_{\Lambda}\mat{A}_{\Lambda}\bigr)}\bm{1}_J$ and $\opt{\bm{\lambda}}=\bm{0}$. Taking the uniqueness of the solution into consideration, we can conclude that the unique minimizer of~\eqref{cvx:NLasso_Lambda} is given by 
\begin{IEEEeqnarray}[]{c}
	\opt{\bm{v}}_{\Lambda} = \mat{A}_{\Lambda}^\dagger\bm{y} - \gamma\inv{\bigl(\tr{\mat{A}}_{\Lambda}\mat{A}_{\Lambda}\bigr)}\bm{1}_J \succ \bm{0}.\label{eq:sol_NLasso_specific}
\end{IEEEeqnarray}
By manipulating the inequality~\eqref{eq:APMRC_NLasso:NSCC} and substituting~\eqref{eq:sol_NLasso_specific}, we have
\begin{IEEEeqnarray*}{r,r,c,l}
	&\tr{(\bm{y} - \mat{A}_\Lambda\mat{A}^\dagger_\Lambda\bm{y})}\bm{a}_j &<& \gamma(1-\tr{\bm{1}}_J\mat{A}_\Lambda^\dagger\bm{a}_j)\\
	\Leftrightarrow & \tr{\bigl(\bm{y} - \mat{A}_\Lambda\mat{A}^\dagger_\Lambda\bm{y}+\gamma\tr{(\mat{A}_\Lambda^\dagger)}\bm{1}_J\bigr)}\bm{a}_j &<& \gamma \\
	\Leftrightarrow &\tr{\bigl\{\bm{y} - \mat{A}_\Lambda\bigl(\mat{A}^\dagger_\Lambda\bm{y}-\gamma\inv{(\tr{\mat{A}_\Lambda}\mat{A}_\Lambda)}\bm{1}_J\bigr)\bigr\}}\bm{a}_j &<& \gamma \\
	\Leftrightarrow &\tr{\bigl( \bm{y} - \mat{A}_\Lambda \opt{\bm{v}}_\Lambda \bigr)}\bm{a}_j &<& \gamma \quad (\because\eqref{eq:sol_NLasso_specific})
\end{IEEEeqnarray*}
for all $j \in \stcomp{\Lambda}$. Directly applying Theorem~\ref{thm:APMRC_NLasso_base}, we can assert that $\Lambda$ contains the support of $\opt{\bm{x}}$. Furthermore, since all the elements of the minimizer~\eqref{eq:sol_NLasso_specific} are greater than zero, $\Lambda$ is the support of $\opt{\bm{x}}$. This completes the proof.

\hfill \IEEEQEDclosed

\subsection{Simplified sufficient conditions for NLasso}
\label{sec:SMRCs}
Although these conditions are quite simple, they are still more elaborate than those provided by Tropp~\cite{Tropp2006TIT} for general lasso. We will further simplify and introduce two sufficient conditions in this section.  To start with, we define the positive exact recovery coefficient ($\mathrm{PERC}$) by
\begin{IEEEeqnarray*}[]{c}
	\mathrm{PERC}(\Lambda) := \min _{j\in \stcomp{\Lambda}} {\mathrm{PSC}(\Lambda;j)}.
\end{IEEEeqnarray*}
We call this $\mathrm{PERC}$ because this is considered as the positive counterpart of the $\mathrm{ERC}$, and $\mathrm{PERC}$ is interpreted as the minimum of the right hand side in the NSCC~\eqref{eq:APMRC_NLasso:NSCC}. Note that $(1-\mathrm{PERC})$ is equivalent to non-negative irrepresentable constant of~\cite{Slawski2013}.
We can also take the maximum on the left hand side of the inequality~\eqref{eq:APMRC_NLasso:NSCC} after concatenating all $\bm{a}_j$, and then a modified sufficient condition for the multiple NSCCs is written as
\begin{IEEEeqnarray*}[]{c}
	\max_{j \in \stcomp{\Lambda}} {\tr{\bm{a}_j}\mat{P}^{\perp}_{\Lambda}\bm{y}} < \gamma\mathrm{PERC}(\Lambda)\IEEEyesnumber\label{eq:NSCC_PERC-Max}.
\end{IEEEeqnarray*}
We refer to this condition~\eqref{eq:NSCC_PERC-Max} as PERC-Max. Using this, we can derive the next corollary.
\vspace{2pt}
\begin{corollary}[PERC-Max MRC for NLasso]\label{cor:SMRC_NLasso_PERC-Max}
	Under the assumptions of Theorem~\ref{thm:APMRC_NLasso}, the support of $\opt{\bm{x}}$, $\mathrm{supp}(\opt{\bm{x}})$, is equal to $\Lambda$ if the two conditions, MCC~\eqref{eq:APMRC_NLasso:MCC} and PERC-Max~\eqref{eq:NSCC_PERC-Max}, hold. Furthermore, the minimizer $\opt{\bm{x}}$ is also given as in Theorem~\ref{thm:APMRC_NLasso}.
\end{corollary}
\vspace{2pt}
We can still introduce another NSCC condition that is more strict than Corollary~\ref{cor:SMRC_NLasso_PERC-Max} but more relaxed than Theorem~\ref{thm:SMRC_NLasso_ERC-based} by taking the absolute maximum value on the left side of~\eqref{eq:NSCC_PERC-Max},
\begin{IEEEeqnarray*}[]{c}
	\ \| \tr{\mat{A}}\mat{P}^{\perp}_{\Lambda}\bm{y}\|_\infty < \gamma\mathrm{PERC}(\Lambda)\IEEEyesnumber\label{eq:NSCC_PERC-AMax},
\end{IEEEeqnarray*}
We refer this condition as PERC-absolute Max (PERC-AMax), leading to the next corollary.
\vspace{2pt}
\begin{corollary}[PERC-AMax MRC for NLasso]\label{cor:SMRC_NLasso_PERC-AMax}
	Under the assumptions of Theorem~\ref{thm:APMRC_NLasso}, the support of $\opt{\bm{x}}$, $\mathrm{supp}(\opt{\bm{x}})$, is equal to $\Lambda$ if the two conditions, MCC~\eqref{eq:APMRC_NLasso:MCC} and PERC-AMax~\eqref{eq:NSCC_PERC-AMax}, hold. Furthermore, the minimizer $\opt{\bm{x}}$ is also given as in Theorem~\ref{thm:APMRC_NLasso}.
\end{corollary}
\vspace{2pt}
PERC-AMax is more demanding than PERC-Max. Additionally, it provides a broader guarantee than Theorem~\ref{thm:SMRC_NLasso_ERC-based}, which assumes the linear model~\eqref{eq:lmm} and requires not only the support of the true coefficient vector $\bm{x}^{\text{true}}$ but also its values and the error term in advance, while Corollary~\ref{cor:SMRC_NLasso_PERC-AMax} can be applied to arbitrary vectors $\bm{y}$ and index sets $\Lambda$.

\section{Application to hyperspectral unmixing}
\label{sec:application}
Hyperspectral imagers collect electromagnetic radiation over the Visible and Near Infrared (VNIR) to Short Wave Infrared (SWIR) region (300-2600nm) with hundreds of narrow contiguous bands. Each pixel position of a hyperspectral image (HSI) is associated with a {\em spectrum} or {\em spectral signature}, an array of dimension equal to the number of bands, which is used by practitioners to reveal the compositional characteristics of targets in a variety of applications. 

One of the tasks routinely performed in hyperspectral imaging is spectral unmixing. Unmixing is a process to decompose an observed spectrum into pure constituent signatures, which are usually called {\em endmembers}, associated with their fractional abundances \cite{Bioucas-Dias2012JSTARS},\cite{Heylen2014}. The linear model~\eqref{eq:lmm} has been widely used in unmixing, where $\bm{y}$ represents the observed spectrum, $\mat{A}$ is the dictionary matrix with atoms corresponding to endmember spectra, and the coefficient vector $\bm{x}$ is interpreted as a fractional abundance vector. 

Recently, sparse unmixing~\cite{Iordache2011TGRS} has been proposed for hyperspectral unmixing tasks where one is given a large collection --- the {\em spectral library} --- of pure spectral signatures to be used as candidate endmembers. 
The first goal for unmixing is to correctly identify the library spectra that are combined to form the observed spectrum. One typically expects that only a few endmembers in the collection are involved in the observation, as the number of materials occupying the region subtended by a pixel is small in most scenarios. Motivated by this observation, sparse unmixing~\cite{Iordache2011TGRS} employs NLasso to detect endmembers and estimate abundances for hyperspectral images using a large library of laboratory spectra. 

Nonetheless, sparse unmixing faces several limitations in practical applications. An observed spectrum could be composed by a {\em nonlinear} mixture of endmembers. The atoms in the library might not match exactly the image endmembers (typical examples are spectra of the same material acquired at different conditions). Moreover, the library spectra are usually highly correlated, which intuitively seems undesirable for the non-negative sparse modeling. 

Given such complications, we propose the theorems derived in this paper as a way to assess the performance of NLasso in hyperspectral unmixing. Since we have not restricted the definition of the ``error" term $\bm{e}$ in the linear model equation \eqref{cvx:linmod}, it could accommodate any deviation from linearity, such as nonlinear mixing or spectral distortions. 

We test the performance of NLasso in unmixing a real hyperspectral image of an oil painting. In particular, we are interested in assessing the performance of the sparse modeling approach in identifying the endmembers involved in each pixel of the HSI. This example presents the typical complications of unmixing problems: the artist creates the colors in the painting by mechanically mixing the paints (nonlinear mixing) and adding water (nonlinear distortion of all pixel spectra). Further nonlinearities stem from the different density and thickness of the paint in different regions.  We first describe the creation of training data including endmembers and their true abundance maps. We then assess the ability of the MRCs to predict the outcome of numerical computations using NLasso.

\subsection{Data set}
\begin{figure}[!t]
	\centering
	\includegraphics[]{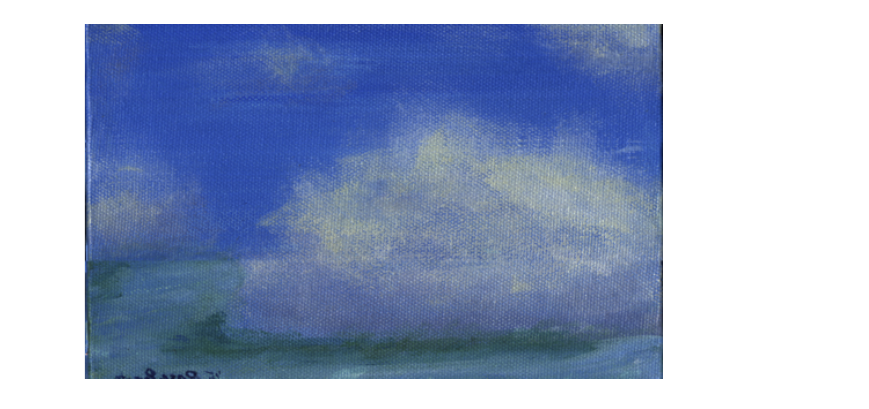}%
	\hfill
	\caption{$500 \times 308$ pseudo RGB color image of a hyperspectral image of the oil painting. R: band 125 (601nm), G: band 86 (536nm), B: band  49 (471nm).}
	\label{fig:painting_rgb}
\end{figure}

The data set used for this experiment is a hyperspectral image (HSI) of an oil painting acquired by a Micro-Hyperspec\textsuperscript{\textregistered} VNIR imaging sensor (E-Series).\footnote{Micro-Hyperspec\textsuperscript{\textregistered} is a registered trademark of Headwall Photonics, Inc.} The imager captures spectral information in the 400-1000nm wavelength region with 370 bands at 1.6nm intervals. The image was acquired on Hyperspec Starter Kit manufactured by Headwall Inc. Since the painting was larger than the field of view of the camera, 
we first separated the whole area into three strip regions, acquired the images of them separately, and stitched them together. The size of the resulting hyperspectral image is $3347\!\times\!1233\!\times\!370$, where the first two dimensions are the number of pixels in the spatial directions and the last dimension records the number of spectral bands. The image was converted to reflectance using a Spectralon\textsuperscript{\textregistered} reference.\footnote{Spectralon\textsuperscript{\textregistered} is a registered trademark of Labsphere, Inc.} The image was then  spatially downsampled to $500\!\times\!308$ by averaging $4\!\times\!4$ neighboring pixels in order to improve its signal-to-noise ratio. 
Figure~\ref{fig:painting_rgb} shows a pseudo RGB rendition of the HSI.

\begin{figure}[!t]
	\centering
	\includegraphics[]{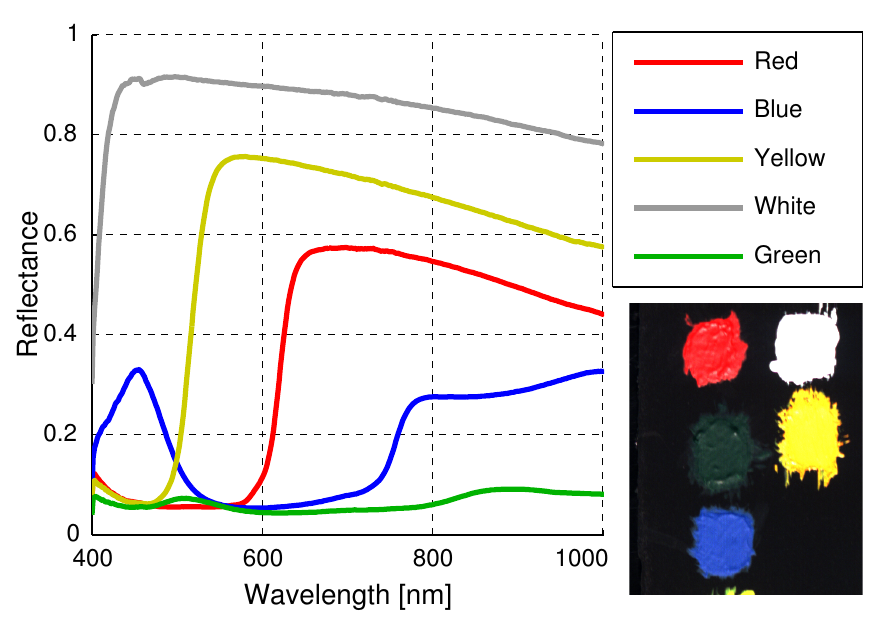}
	\caption{Averaged endmember spectra. The image on the lower right is a pseudo RGB color image of a hyperspectral image of endmembers.}
	\label{fig:endmember_spectra}
\end{figure}

The artist used acrylic paint in five distinct colors: red, blue, yellow, white, and green. We acquired an HSI of the pure colors, which is shown on the lower right side in Figure~\ref{fig:endmember_spectra}. The averaged spectra of the five colored areas can be considered as the endmembers of the pixels depicting the painting: they are shown in Figure~\ref{fig:endmember_spectra}, and we use them to construct our spectral library. 

\begin{figure}[!t]
	\centering
	\includegraphics[]{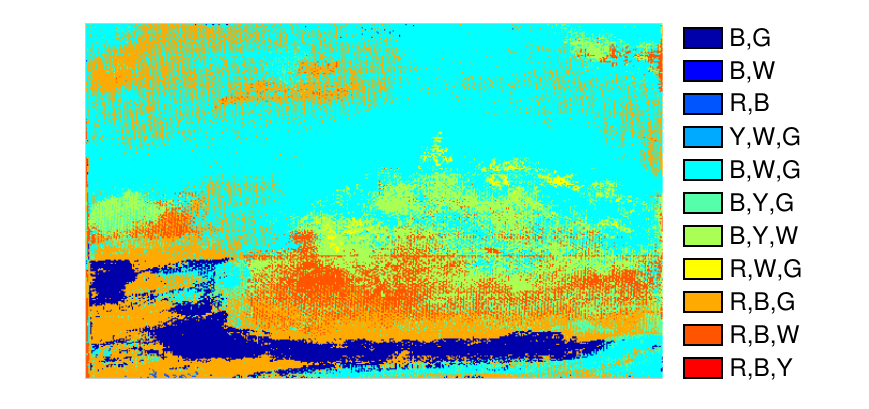}%
	\caption{The true distribution of the mixtures. The letters R, B, Y, W, and G are the first letters of the pure colors.\label{fig:true_mixtypes}}
\end{figure}

The artist did not provide in advance the information about true distribution of the endmembers for each pixel. We will use the term ``ground truth", borrowed from hyperspectral remote sensing, for such map. This information is required in order to assess the performance of the unmixing algorithm. Since the artist used 2 or 3 colors at each location (single-endmember pixels are not present) we could generate the ground truth by solving the following minimization problem for each pixel
\begin{IEEEeqnarray}{l}\label{cvx:NLasso_3}
	\begin{IEEEeqnarraybox}[][c]{l'l}
		\underset{\bm{x}}{\text{minimize}}
		& \frac{1}{2}\|\bm{y}-\mat{A}\bm{x}\|_2^2\\
		\text{subject to}
		& \bm{x} \succeq \bm{0}, \left\| \bm{x} \right\|_0 \le 3,
	\end{IEEEeqnarraybox}
\end{IEEEeqnarray}
where $\bm{y}\in \mathbb{R}^{370}$ is the spectral signal of each pixel, $\mat{A} \in \mathbb{R}^{370\times 5}$ is the matrix of the spectral library of five colors, and $\bm{x}\in \mathbb{R}^{5}$ is the fractional abundance vector. In practice, we found the solution by conducting least square minimizations for all possible combinations of less than or equal to three endmembers and we selected as the ground truth the combination with the smallest number of endmembers among the ones that achieved the least error. Figure~\ref{fig:true_mixtypes} shows the ground truth at each pixel. The distribution map in Figure~\ref{fig:true_mixtypes} shows that not all the pixels are assigned to mixtures of three endmembers; some pixels are to mixtures of only two endmembers, which means that any other third endmember cannot help reduce the residual for these pixels. We could also interpret this fact from our theorem. The minimization problem~\eqref{cvx:NLasso_3} can be converted to NLasso~\eqref{cvx:NLasso} with an appropriate trade-off parameter $\gamma$. We conjecture that a combination of only two endmembers suffices to meet the APMRC in Theorem~\ref{thm:APMRC_NLasso}.

\begin{figure}[!t]
	\centering
	\includegraphics[]{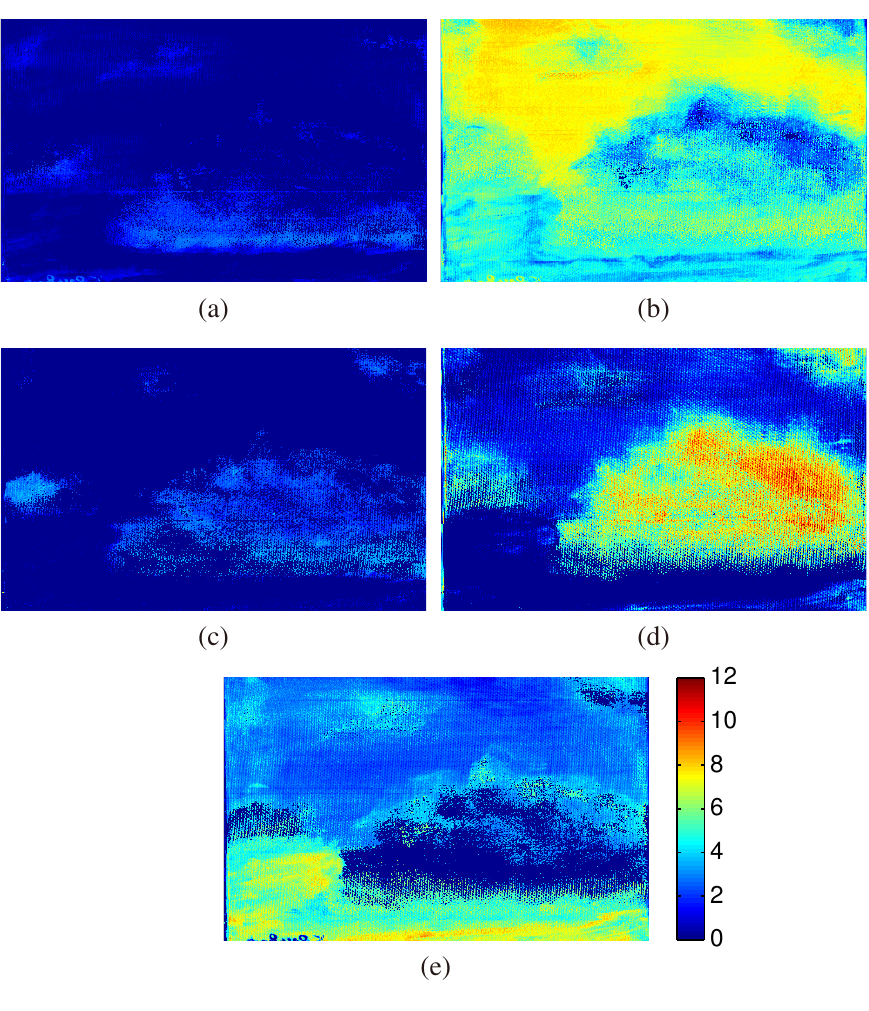}%
	\caption{The true abundance maps in the painting image: (a) red, (b) blue, (c) yellow, (d) white, and (e) green.\label{fig:true_abus}}
\end{figure}

We note that although the minimization problem above also produces abundance values for each endmember at each pixel location, we are discarding such information as we are interested in endmember retrieval performance. Furthermore, while we were able to confirm with the artist that the retrieved distribution is largely accurate, a similar assessment would not be possible for the abundances. We nevertheless report the abundances for all colors in Figure~\ref{fig:true_abus} for the sake of completeness and to show that they show reasonable effects. From Figure~\ref{fig:true_abus}, one can see that the white color is dominant around the clouds in Figure~\ref{fig:painting_rgb}, the sky is mainly painted in white and blue, and the grass that can be found around the bottom in Figure~\ref{fig:painting_rgb} consists of mainly green.
\subsection{Theorem validation}
In this section, we evaluate the predictive power of the MRCs. The performance of NLasso is predicted by four MRCs: Theorem~\ref{thm:SMRC_NLasso_ERC-based} (ERC-based MRC), Theorem~\ref{thm:APMRC_NLasso} (APMRC), Corollary~\ref{cor:SMRC_NLasso_PERC-Max} (PERC-Max MRC), and Corollary~\ref{cor:SMRC_NLasso_PERC-AMax} (PERC-AMax MRC). The algorithm we employed to produce numerical solutions for NLasso is the sparse unmixing by variable splitting and augmented Lagrangian (SUnSAL)~\cite{Bioucas-Dias2010}. We are interested in determining how well the predictions of the MRCs match the result obtained by SUnSAL.

Table~\ref{table:resultpainting} shows the number of pixels that satisfy each of MRCs hold for a different value of $\gamma = 0.2, 0.1, 0.05$. For each value of $\gamma$ we display the confusion matrix between the prediction by each MRC and the SUnSAL result. In each row, the label ``True" refers to the points fulfilling the specific condition and ``False" the ones violating it. Similarly, in each column, the label "Correct" (``Incorrect") refers to points for which SUnSAL correctly identifies (fails to identify) the endmembers. 

For all the MRCs, the values in the True-Incorrect cells are all zeros, indicating that if the MRCs hold, the SUnSAL always succeeds. This is a confirmation for the sufficiency of all the MRCs in all the theorems.  

One remarkable fact is that the ``False-Correct" cell of the APMRC is always zero, which means the APMRC is always true when SUnSAL succeeds in detecting the correct endmembers, confirming the necessity of the APMRC condition in practical settings. In contrast, there are non-zero values in the False-Correct cells for the other MRCs, and the values increase as the conditions become increasingly strict. 

None of the pixels satisfy the strict conditions required by the ERC-based MRC. It is worth noting that the application we chose tests the limits of the theory of sparse recovery in at least two ways. First, hyperspectral mixing processes can deviate significantly from the linear model; second, spectra of different endmembers are very correlated. In our view, this observation supports the utility of the APMRC, PERC-Max MRC and PERC-AMax MRC as prediction tools for signal spaces in which previously proposed metrics would not be applicable. \textcolor{red}{}
\begin{table*}[t]
	\renewcommand{\arraystretch}{1.4}
	\setlength{\extrarowheight}{1.5pt}
	\caption{Performance of MRCs on the painting data}
	\label{table:resultpainting}
	\centering	
	\begin{tabular}[c]{|l|c|*{6}{C{40pt}|}}
		\cline{3-8}
		\multicolumn{2}{c|}{} & \multicolumn{2}{c|}{$\gamma=0.2$} & \multicolumn{2}{c|}{$\gamma=0.1$} & \multicolumn{2}{c|}{$\gamma=0.05$} \tabularnewline
		\cline{3-8}
		\multicolumn{2}{c|}{}         & \multicolumn{2}{c|}{SUnSAL retrieval} & \multicolumn{2}{c|}{SUnSAL retrieval} & \multicolumn{2}{c|}{SUnSAL retrieval} \tabularnewline
		\cline{3-8}
		\multicolumn{2}{c|}{}         & Correct & Incorrect & Correct & Incorrect & Correct & Incorrect \tabularnewline
		\cline{1-8}
		\multirow{2}{*}{\parbox[c]{75pt}{Thm.~\ref{thm:APMRC_NLasso}\\(APMRC)}} & True  & 56718 [pts]  & 0 & 70256  & 0 & 64459  & 0 \tabularnewline
		\cline{2-8}
		& False & 0   & 97282 & 0 & 83744 & 0 & 89541 \tabularnewline
		\cline{1-8}
		\multirow{2}{*}{\parbox[c]{75pt}{Cor.~\ref{cor:SMRC_NLasso_PERC-Max}\\(PERC-Max $\mathrm{MRC}$)}} & True  & 56053   & 0 & 67252 & 0 & 62547 & 0\tabularnewline
		\cline{2-8}
		& False & 665   &  97282 &3004 & 83744 & 1912 & 89541\tabularnewline
		\cline{1-8}
		\multirow{2}{*}{\parbox[c]{75pt}{Cor.~\ref{cor:SMRC_NLasso_PERC-AMax}\\(PERC-AMax $\mathrm{MRC}$)}} & True  & 46349   & 0 & 53122 & 0 & 32361 & 0\tabularnewline
		\cline{2-8}
		& False & 10369   &  97282  & 17134 & 83744 & 32098 & 89541\tabularnewline
		\cline{1-8}
		\multirow{2}{*}{\parbox[c]{75pt}{Thm.~\ref{thm:SMRC_NLasso_ERC-based}\\(ERC-based $\mathrm{MRC}$)}} & True  & 0   & 0 & 0 & 0 & 0 & 0\tabularnewline
		\cline{2-8}
		& False & 56718   &  97282  & 70256 & 83744 & 64459 & 89541\tabularnewline
		\cline{1-8}
	\end{tabular}
\end{table*}

\begin{figure}[!t]
	\centering
	\includegraphics[]{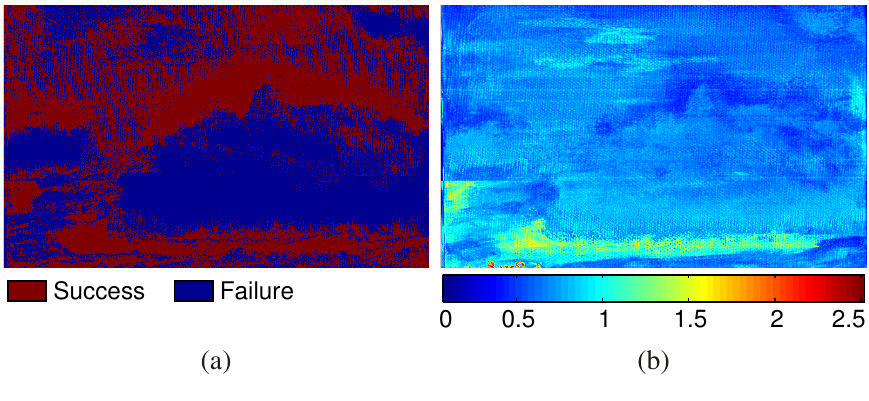}%
	\caption{(a) The distribution of correct model recovery of NLasso and (b) the distribution of the deviation from the linear model.}\label{fig:painting_detection_dev}
\end{figure}

\begin{figure}[!t]
	\centering
	\includegraphics[]{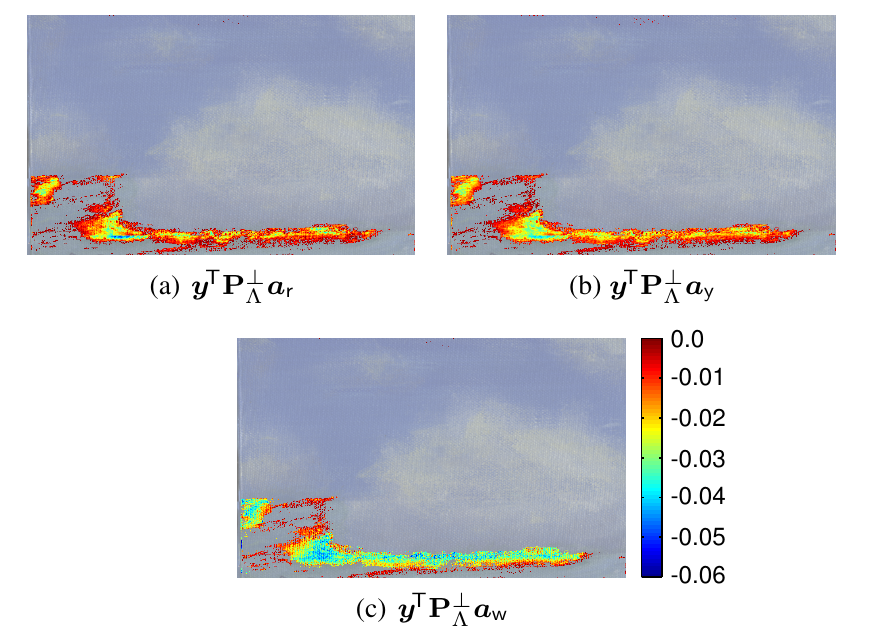}%
	\caption{Mappings of the values of the left hand side of the NSCC for each atom in $\stcomp{\Lambda}$. The letters, $\mathsf{r}$, $\mathsf{y}$, and $\mathsf{w}$, means the first letter of the three colors: red, yellow, and white.\label{fig:anal_detail}}
\end{figure}

Figure~\ref{fig:painting_detection_dev} shows an interesting behavior in the performance of NLasso.
In Figure~\ref{fig:painting_detection_dev}(a), the pixels where NLasso succeeds at identifying endmembers are shown in red, while failures are shown in blue. Figure~\ref{fig:painting_detection_dev}(b) shows the distribution of the residual given by the optimum value of~\eqref{cvx:NLasso_3}. This residual is interpreted as the deviation from linearity; therefore, one may expect that a large deviation is linked to failure of NLasso, but this intuition does not bear out in practice. Deviations seem to be relatively high in the horizontal belt near the bottom of the painting, but NLasso is able to detect the correct endmembers in that area. The reason for this phenomenon is explained in~Figure~\ref{fig:anal_detail}. In this figure, we focus only on the region classified as the mixture of blue and green that mostly overlaps with the horizontally belted region. Figure~\ref{fig:anal_detail} shows the distribution of the values $\tr{\bm{y}}\mat{P}^{\perp}_{\Lambda}\bm{a}_j$ where $j$ indexes the red, yellow, and white colors, which are associated with the subscripts $\mathsf{r}$, $\mathsf{y}$, and $\mathsf{w}$, respectively. The values in the region of interest in Figures~\ref{fig:anal_detail} are always negative, while their corresponding $\mathrm{PSC}(\Lambda;j)$ are always positive ($\mathrm{PSC}(\Lambda;\mathsf{r})=0.141$, $\mathrm{PSC}(\Lambda;\mathsf{y})=0.151$, and $\mathrm{PSC}(\Lambda;\mathsf{w})=0.061$), meaning the NSCC conditions for the APMRC, PERC-Max MRC, PERC-AMax MRC are always true. Even when the deviation is large, NLasso is successful if the residual is negatively correlated with the signatures of all the other endmembers and $\mathrm{PSC}$ is positive and vice versa. Although this does not always happen, this specific example demonstrates that the direction of the deviation affects the performance of NLasso.


One final observation about Figure~\ref{fig:painting_detection_dev}(b) is that the region of maximum deviation from nonlinearity (green to yellow) corresponds with a region to which the artist confirmed having applied more than one layer of paint (this is called {\em pentimento} in art jargon). This is consistent with our understanding of the radiative transfer aspects of unmixing, e.g \cite{Heylen2014} and suggests the potential utility of unmixing techniques to aspects of the artist painting style in addition to identifying the different pigments.

\section{Conclusion}
\label{sec:conclusion}
In this paper, we explored several recovery conditions that guarantee the correct identification of endmembers by the non-negative sparse modeling for mixed signals exhibiting deviations from linearity. Those conditions reveal an interesting property of NLasso, which is expressed by two conditions: minimum coefficient condition and nonlinearity vs subset coherence condition. In particular, we derived an almost perfect condition that can exactly predict the performance of NLasso in a practical sense. The exactness was inferred from mathematical inspection and further verified through experiments. These conditions are proven to be useful for analyzing the performance of numerical solutions of NLasso. 

\appendix[proof of Theorems~\ref{thm:APMRC_NLasso_base} and~\ref{thm:APMRC_NLasso_base_eq}]
\label{appdx:proof:APMRC_NLasso_base}
In this appendix, we prove Theorems~\ref{thm:APMRC_NLasso_base} and~\ref{thm:APMRC_NLasso_base_eq} together because they share assumptions. 

\noindent Recall the NLasso~\eqref{cvx:NLasso2} is expressed as:
\begin{IEEEeqnarray*}{l}
	\begin{IEEEeqnarraybox}[][c]{l'l}
		\underset{\bm{x}}{\text{minimize}}
		& \frac{1}{2}\|\bm{y}-\mat{A}\bm{x}\|_2^2 + \gamma \tr{\bm{1}_N} \bm{x} \\
		\text{subject to}
		& \bm{x} \succeq \bm{0}
	\end{IEEEeqnarraybox}
\end{IEEEeqnarray*}
where $\mat{A} \in \mathbb{R}^{L\times N}$ is a dictionary matrix and the $\bm{y} \in \mathbb{R}^L$ is an observation signal. We prove the next two statements:
\begin{IEEEeqnarray*}[]{t't}
	(Theorem~\ref{thm:APMRC_NLasso_base}) & \eqref{eq:APMRC_NLasso_base} $\Rightarrow$ $\supp{\opt{\bm{x}}} \subseteq \Lambda$ \IEEEyesnumber\label{statement1}\\
	(Theorem~\ref{thm:APMRC_NLasso_base_eq}) & $\supp{\opt{\bm{x}}} \subseteq \Lambda$ $\Rightarrow$ \eqref{eq:APMRC_NLasso_base_eq} \IEEEyesnumber\label{statement2}
\end{IEEEeqnarray*}
where $\opt{\bm{x}}$ is a solution to NLasso. 

Before proceeding with the proof, we first transform the event $\supp{\opt{\bm{x}}} \subseteq \Lambda$ into an equivalent form. More specifically, $\supp{\opt{\bm{x}}} \subseteq \Lambda$ means that the solution of NLasso is equivalent to a solution of the restricted problem~\eqref{cvx:NLasso_Lambda}, $\opt{\bm{v}}_\Lambda$, with appropriate zero-padding; this can be written as $\mat{A}\opt{\bm{x}} = \mat{A}_{\Lambda}\opt{\bm{v}}_{\Lambda}$, 
where $\opt{\bm{v}}_\Lambda$ is an optimal solution of the restricted NLasso. We rewrite the problem as
\begin{IEEEeqnarray}{c}
	\begin{IEEEeqnarraybox}[][c]{r'l}
		\underset{\bm{v}_\Lambda}{\text{minimize}}
		& \frac{1}{2}\|\bm{y}-\mat{A}_\Lambda\bm{v}_\Lambda\|_2^2 + \gamma\tr{\bm{1}}_J\bm{v}_\Lambda\\
		\text{subject to}
		& \bm{v}_\Lambda \succeq \bm{0}.
	\end{IEEEeqnarraybox}
	\label{cvx:NLasso_Lambda2}
\end{IEEEeqnarray}
Thus, we have $\mathrm{supp}(\opt{\bm{x}})\subseteq \Lambda$ if and only if the following inequality holds:
\ifCLASSOPTIONdraftcls
\begin{IEEEeqnarray*}{c}
	\begin{IEEEeqnarraybox*}[][c]{r,c,l}
		\frac{1}{2}\|\bm{y}-\mat{A}\opt{\bm{x}}\|_2^2 + \gamma \tr{\bm{1}}_N\opt{\bm{x}} &<& \frac{1}{2}\|\bm{y}-\mat{A}(\opt{\bm{x}}+\Delta\bm{x})\|_2^2 + \gamma \tr{\bm{1}}_N(\opt{\bm{x}}+\Delta\bm{x})
	\end{IEEEeqnarraybox*}
	\IEEEyesnumber\label{eq:proof:PMRC_NLasso_Lambda:primitiveperturbIneq}
\end{IEEEeqnarray*}
\else
\begin{IEEEeqnarray*}{c}
	\begin{IEEEeqnarraybox*}[][c]{r,c,l}
		\frac{1}{2}\|\bm{y}-\mat{A}\opt{\bm{x}}\|_2^2 + \gamma \tr{\bm{1}}_N\opt{\bm{x}} &<& \frac{1}{2}\|\bm{y}-\mat{A}(\opt{\bm{x}}+\Delta\bm{x})\|_2^2 \\ && \hfill + \gamma \tr{\bm{1}}_N(\opt{\bm{x}}+\Delta\bm{x})
	\end{IEEEeqnarraybox*}
	\IEEEyesnumber\label{eq:proof:PMRC_NLasso_Lambda:primitiveperturbIneq}
\end{IEEEeqnarray*}
\fi
for every $\Delta \bm{x}$ such that $0 < \Delta x_j$ for any $j \in \stcomp{\Lambda}$ and $\bm{x}+\Delta \bm{x}\succeq\bm{0}$ where $\Delta x_j$ is the $j^\textrm{th}$ element of $\bm{x}$. In other words, the minimum cost achieved by the solution of the restricted problem is less than any cost achieved by another $\bm{x}$ that involves an atom outside $\Lambda$. 
Let the exact support of $\opt{\bm{x}}_\Lambda$ be $\Gamma \subseteq \Lambda$ $(|\Gamma| = M \le J)$. Then the inequality $\bm{x}+\Delta \bm{x}\succeq\bm{0}$ is expressed as:
\begin{IEEEeqnarray*}[]{c}
	\begin{IEEEeqnarraybox*}[][c]{-t'l}
		1) & -x_j \le \Delta x_j \text{ for all } j \in \Gamma \\ 
		2) & 0 \le \Delta x_j \text{ for all } j \in \stcomp{\Gamma},
	\end{IEEEeqnarraybox*}\IEEEyesnumber\label{eq:proof:PMRC_NLasso_Lambda:pert2}
\end{IEEEeqnarray*}
which defines a region of interest for the vector $\Delta \bm{x}$.
By canceling terms common to both sides, the inequality~\eqref{eq:proof:PMRC_NLasso_Lambda:primitiveperturbIneq} is transformed into
\ifCLASSOPTIONdraftcls
	\begin{IEEEeqnarray*}{l}
		\frac{1}{2}\left\| \mat{A}(\Delta \bm{x})\right\| _2^2
		+ 
		\gamma \tr{\bm{1}}_N(\Delta \bm{x})
		-
		\tr{\left( \bm{y} - \mat{A}_{\Lambda}\opt{\bm{v}}_{\Lambda} \right) }\! \mat{A}(\Delta \bm{x})> 0 \\ 
		\Leftrightarrow 
		\frac{1}{2}\left\| \mat{A}(\Delta \bm{x})\right\| _2^2
		+ \sum_{j \in \Omega}{\Delta x_j\bigl(\gamma-\tr{\bm{a}_j}(\bm{y}-\mat{A}_\Lambda\opt{\bm{v}}_\Lambda) \bigl) } > 0, \IEEEyesnumber\label{eq:proof:PMRC_NLasso_Lambda:simpPert}
	\end{IEEEeqnarray*}
\else
	\begin{IEEEeqnarray*}{,l}
		\frac{1}{2}\left\| \mat{A}(\Delta \bm{x})\right\| _2^2
		+ 
		\gamma \tr{\bm{1}}_N(\Delta \bm{x})
		-
		\tr{\left( \bm{y} - \mat{A}_{\Lambda}\opt{\bm{v}}_{\Lambda} \right) }\! \mat{A}(\Delta \bm{x})> 0 \\ 
		\Leftrightarrow 
		\frac{1}{2}\left\| \mat{A}(\Delta \bm{x})\right\| _2^2
		+ \sum_{j \in \Omega}{\Delta x_j\bigl(\gamma-\tr{\bm{a}_j}(\bm{y}-\mat{A}_\Lambda\opt{\bm{v}}_\Lambda) \bigl) } > 0, \IEEEyesnumber\label{eq:proof:PMRC_NLasso_Lambda:simpPert}
	\end{IEEEeqnarray*}
\fi
where $\Omega = \{1,2,\ldots,N\}$ is the whole column index set as defined in Section~\ref{sec:mathnotation}. Next, we explore a property of $\opt{\bm{v}}_\Lambda$. The Lagrangian of the restricted NLasso \eqref{cvx:NLasso_Lambda2} above is given by
\begin{IEEEeqnarray*}[]{c}
	L_\Lambda (\bm{v}_\Lambda,\bm{\lambda}_\Lambda) =  \frac{1}{2}\|\bm{y}-\mat{A}_\Lambda\bm{v}_\Lambda\|_2^2 + \gamma\tr{\bm{1}}_J\bm{v}_\Lambda - \tr{\bm{\lambda}_\Lambda}\bm{v}_\Lambda,
\end{IEEEeqnarray*}
where $\bm{\lambda}_\Lambda$ is a vector of Lagrangian multipliers. From the KKT condition in Theorem 28.3~\cite[p. 281]{ConvexAnalysis}, $\bm{v}_\Lambda = \opt{\bm{v}}_\Lambda$ and $\bm{\lambda}=\opt{\bm{\lambda}}$ become a minimizer and a Kuhn-Tucker vector, respectively, if and only if the following three conditions hold:
\ifCLASSOPTIONdraftcls
\begin{IEEEeqnarray*}[]{'s'L}
	1) & \opt{\bm{v}}_\Lambda, \opt{\bm{\lambda}}_\Lambda \succeq \bm{0} \IEEEyesnumber\IEEEyessubnumber\label{eq:KKTcnd1_PL-NLass_Lambda}\\
	2) & \opt{\bm{\lambda}}_\Lambda(n)\opt{\bm{v}}_\Lambda(n) = 0 \text{ for all } n \IEEEyessubnumber\label{eq:KKTcnd2_NLasso_Lambda}\\
	3) & \bm{0} = \partial L_\Lambda(\opt{\bm{v}}_\Lambda, \opt{\bm{\lambda}}_\Lambda)  / \partial \bm{v}_\Lambda|_{\bm{v}_\Lambda=\opt{\bm{v}}_\Lambda}\IEEEyessubnumber\label{eq:KKTcnd3_NLasso_Lambda}.
\end{IEEEeqnarray*}
\else
\begin{IEEEeqnarray*}[]{-s'L}
	\;\;1) & \opt{\bm{v}}_\Lambda, \opt{\bm{\lambda}}_\Lambda \succeq \bm{0} \IEEEyesnumber\IEEEyessubnumber\label{eq:KKTcnd1_PL-NLass_Lambda}\\
	\;\;2) & \opt{\bm{\lambda}}_\Lambda(n)\opt{\bm{v}}_\Lambda(n) = 0 \text{ for all } n \IEEEyessubnumber\label{eq:KKTcnd2_NLasso_Lambda}\\
	\;\;3) & \bm{0} = \partial L_\Lambda(\opt{\bm{v}}_\Lambda, \opt{\bm{\lambda}}_\Lambda)  / \partial \bm{v}_\Lambda|_{\bm{v}_\Lambda=\opt{\bm{v}}_\Lambda}\IEEEyessubnumber\label{eq:KKTcnd3_NLasso_Lambda}.
\end{IEEEeqnarray*}
\fi
The third KKT condition~\eqref{eq:KKTcnd3_NLasso_Lambda} is equivalently expressed as:
\begin{IEEEeqnarray*}[]{r'l}
	& \tr{\mat{A}}_\Lambda (\mat{A}_\Lambda \opt{\bm{v}}_\Lambda -\bm{y}) + \gamma \bm{1}_J - \opt{\bm{\lambda}}_\Lambda = \bm{0} \\
	\Leftrightarrow & \gamma -\tr{\bm{a}}_j (\bm{y}-\mat{A}_\Lambda \opt{\bm{v}}_\Lambda ) = \opt{\bm{\lambda}}_\Lambda(j) \quad\text{for all } j\in\Lambda 
\end{IEEEeqnarray*}
For $j \in \Gamma$, we have $\opt{\bm{v}}_\Lambda(j)>0$, leading to $\opt{\bm{\lambda}}_\Lambda(j) = 0$ because of the second KKT condition~\eqref{eq:KKTcnd2_NLasso_Lambda}. For $j \in \Lambda \setminus \Gamma$, we have $\opt{\bm{v}}_\Lambda(j)=0$, leading to $\opt{\bm{\lambda}}_\Lambda(j) \ge 0$. Thus we have
\begin{IEEEeqnarray}[]{c}
	\begin{cases}
		\gamma -\tr{\bm{a}}_j (\bm{y}-\mat{A}_\Lambda \opt{\bm{v}}_\Lambda ) = 0 \text{ for } j \in \Gamma \\
		\gamma -\tr{\bm{a}}_j (\bm{y}-\mat{A}_\Lambda \opt{\bm{v}}_\Lambda ) \ge 0 \text{ for } j \in \Lambda\!\setminus\!\Gamma.
	\end{cases}\IEEEyesnumber\label{eq:proof:PMRC_NLasso_Lambda:kkt3derivativesimple}
\end{IEEEeqnarray}
Considering the conditions above for $\opt{\bm{v}}_\Lambda$, the inequality~\eqref{eq:proof:PMRC_NLasso_Lambda:simpPert} is equivalently transformed into
\ifCLASSOPTIONdraftcls
	\begin{IEEEeqnarray*}{,l}
		\frac{1}{2}\left\| \mat{A}(\Delta \bm{x})\right\| _2^2
		+ \sum_{j \in \Lambda \setminus \Gamma}{\Delta x_j\bigl(\gamma-\tr{\bm{a}_j}(\bm{y}-\mat{A}_\Lambda\opt{\bm{v}}_\Lambda) \bigl) } + \sum_{j \in \stcomp{\Lambda}}{\Delta x_j\bigl(\gamma-\tr{\bm{a}_j}(\bm{y}-\mat{A}_\Lambda\opt{\bm{v}}_\Lambda) \bigl) } > 0, \IEEEyesnumber\label{eq:proof:APMRC_NLasso_base:simpPert2}
	\end{IEEEeqnarray*}
\else
	\begin{IEEEeqnarray*}{,l}
		\frac{1}{2}\left\| \mat{A}(\Delta \bm{x})\right\| _2^2
		+ \sum_{j \in \Lambda \setminus \Gamma}{\Delta x_j\bigl(\gamma-\tr{\bm{a}_j}(\bm{y}-\mat{A}_\Lambda\opt{\bm{v}}_\Lambda) \bigl) } \\ \quad + \sum_{j \in \stcomp{\Lambda}}{\Delta x_j\bigl(\gamma-\tr{\bm{a}_j}(\bm{y}-\mat{A}_\Lambda\opt{\bm{v}}_\Lambda) \bigl) } > 0, \IEEEyesnumber\label{eq:proof:APMRC_NLasso_base:simpPert2}
	\end{IEEEeqnarray*}
\fi
where the second term is always non-negative because of the non-negativity of the two factors (\eqref{eq:proof:PMRC_NLasso_Lambda:pert2}
and \eqref{eq:proof:PMRC_NLasso_Lambda:kkt3derivativesimple}). Summarizing this discussion, $\mathrm{supp}(\opt{\bm{x}})\subseteq\Lambda$ if and only if the inequality~\eqref{eq:proof:APMRC_NLasso_base:simpPert2} holds for all $\Delta \bm{x}$ in the defined region \eqref{eq:proof:PMRC_NLasso_Lambda:pert2}. 

We now prove the statement~\eqref{statement1} for Theorem~\ref{thm:APMRC_NLasso_base}. Given the condition~\eqref{eq:APMRC_NLasso_base} is true, then the summation of the third term on the left side in~\eqref{eq:proof:APMRC_NLasso_base:simpPert2} becomes always non-negative. Furthermore, because we are considering $\Delta \bm{x}$ with a non-zero $j^\textrm{th}$ element for any $j \in \stcomp{\Lambda}$, the third term is always strictly positive. Therefore, \eqref{eq:proof:APMRC_NLasso_base:simpPert2} holds for every $\Delta \bm{x}$ in the defined region~(\ref{eq:proof:PMRC_NLasso_Lambda:pert2}). Because that condition is equivalent to $\mathrm{supp}(\opt{\bm{x}}) \subseteq \Lambda$, the statement~\eqref{statement1} is proven.

Next, we prove the statement~\eqref{statement2} for Theorem~\ref{thm:APMRC_NLasso_base_eq}. We prove this by the principle of contradiction. Assume the inequality~\eqref{eq:proof:APMRC_NLasso_base:simpPert2} is true for every $\Delta \bm{x}$ in the defined region and every solution $\opt{\bm{v}}_\Lambda$. Suppose there exists a solution $\opt{\bm{v}}_\Lambda$ and an index $j' \in \stcomp{\Lambda}$ such that 
\begin{IEEEeqnarray*}{c}
	\tr{(\bm{y}-\mat{A}_\Lambda\opt{\bm{v}}_\Lambda)}\bm{a}_{j'} > \gamma,
\end{IEEEeqnarray*}
which is the opposite of (\ref{eq:APMRC_NLasso_base_eq}). The inequality~\eqref{eq:proof:APMRC_NLasso_base:simpPert2} is true for $\Delta \bm{x}'$ such that only the $j'^{\textrm{th}}$ element is greater than zero and the others are zero. Let such a $\Delta \bm{x}'$ be 
\begin{IEEEeqnarray}[]{c}
	\Delta \bm{x}' = \tr{[0,\,\ldots\,0,\,\Delta x_{j'},\,0,\, \dots\,0]} \quad (\Delta x_{j'}>0).
\end{IEEEeqnarray}	
The inequality~\eqref{eq:proof:APMRC_NLasso_base:simpPert2} then becomes
\ifCLASSOPTIONdraftcls
	\begin{IEEEeqnarray}[]{c}
		\frac{1}{2}\|\bm{a}_{j'}\|_2^2(\Delta x_{j'})^2 - \bigl(\tr{(\bm{y}-\bm{A}_\Lambda\opt{\bm{v}}_\Lambda)}\bm{a}_{j'}-\gamma\bigr)\Delta x_{j'} > 0. \label{eq:aux}
	\end{IEEEeqnarray}
\else
	\begin{IEEEeqnarray}[]{-c}
		\frac{1}{2}\|\bm{a}_{j'}\|_2^2(\Delta x_{j'})^2 - \bigl(\tr{(\bm{y}-\bm{A}_\Lambda\opt{\bm{v}}_\Lambda)}\bm{a}_{j'}-\gamma\bigr)\Delta x_{j'} > 0. \label{eq:aux}
	\end{IEEEeqnarray}
\fi
The left hand side is a quadratic function with regard to a scalar variable $\Delta x_{j'}$. By defining the quadratic equation's coefficients as
\begin{IEEEeqnarray}[]{r,c,l}
	b_{j'} &=& \frac{1}{2}\|\bm{a}_{j'}\|_2^2 > 0 \\
	c_{j'} &=& 	\tr{(\bm{y}-\bm{A}_\Lambda\opt{\bm{v}}_\Lambda)}\bm{a}_{j'}-\gamma> 0,
\end{IEEEeqnarray}
the quadratic inequality (\ref{eq:aux}) becomes
\begin{IEEEeqnarray}[]{c}
	\Delta x_{j'}(b_{j'}\Delta x_{j^\prime} - c_{j'}) > 0.
\end{IEEEeqnarray}
Because both the coefficients $b_{j'}$ and $c_{j'}$ are positive, the left hand side becomes negative for sufficiently small $\Delta x_{j'}$ such that $0<\Delta x_{j'} < c_{j'}/b_{j^\prime}$. Since $\Delta x_{j'}$ can take any positive value, we can say that there exists a $\Delta x_{j'}$ that breaks the inequality~\eqref{eq:proof:APMRC_NLasso_base:simpPert2}. This contradicts to our starting assumption. Thus, by the principle of contradiction, the statement~\eqref{statement2} is proven. 

\hfill \IEEEQEDopen

\section*{Acknowledgments}
We thank Rose Kontak for providing us with the oil painting, and Dr. Bioucas-Dias and Dr. Figueiredo for making Matlab code for SUnSAL available online.

\ifCLASSOPTIONcaptionsoff
  \newpage
\fi

\bibliographystyle{IEEEtran_yuki}


%
%
\end{document}